\documentclass[a4paper,11pt]{article}
\pdfoutput=1 
\usepackage{jinstpub} 

\usepackage{lineno}
\usepackage{subfigure}
\usepackage{comment}

\usepackage{siunitx}

\usepackage{placeins}

\newcommand{\neqcm}{\ensuremath{\mathrm{n}_{\mathrm{eq}}/\mathrm{cm}^2}}

\newcommand\fig{figure}

\newcommand\sect{section}

\newcommand\tab{table}

\graphicspath{{./images/}}

\title{Characterisation of AMS H35 HV-CMOS monolithic active pixel sensor prototypes for HEP applications}

\author[a,1]{S.~Terzo,\note{Corresponding author.}}
\author[b]{M.~Benoit,}
\author[a]{E.~Cavallaro,}
\author[a]{R.~Casanova,}
\author[b]{F.A.~Di~Bello,}
\author[a]{F.~F\"orster,}
\author[a,c]{S.~Grinstein,}
\author[b]{G.~Iacobucci,}
\author[d]{I.~Per\'ic,}
\author[a]{C.~Puigdengoles,}
\author[b]{M.~Vicente~Barrero~Pinto,}
\author[e]{and E.~Vilella~Figueras}

\affiliation[a]{Institut de F\'isica d'Altes Energies (IFAE), The Barcelona Institute of Science and Technology, Edifici CN, UAB campus, 08193 Bellaterra (Barcelona), Spain}
\affiliation[b]{D\'epartement de Physique Nucl\'eaire et Corpusculaire (DPNC), University of Geneva, 24 quai Ernest Ansermet, 1211 Gen\`eve 4, Switzerland}
\affiliation[c]{Instituci\'o Catalana de Recerca i Estudis Avan\c{c}ats (ICREA), Pg. Llu\'is Companys 23, 08010 Barcelona, Spain}
\affiliation[d]{Karlsruher Institut f\"ur Technologie (KIT), Keiserstra{\ss}e 12, 76131 Karlsruhe, Germany}
\affiliation[e]{University of Liverpool, Oliver Lodge, Oxford Street, Liverpool L69 7ZE, United Kingdom}

\emailAdd{Stefano.Terzo@cern.ch}

\abstract{Monolithic Active Pixel Sensors (MAPS) produced in High Voltage CMOS (HV-CMOS) technology are being considered for High Energy Physics applications due to the ease of production and the reduced costs. Such technology is especially appealing when large areas to be covered and material budget are concerned. This is the case of the outermost pixel layers of the future ATLAS tracking detector for the High Luminosity LHC. For experiments at hadron colliders, radiation hardness is a key requirement which is not fulfilled by standard CMOS sensor designs that collect charge by diffusion. This issue has been addressed by depleted active pixel sensors in which electronics are embedded into a large deep implantation ensuring uniform charge collection by drift. Very first small prototypes of hybrid depleted active pixel sensors have already shown a radiation hardness compatible with the ATLAS requirements. Nevertheless, to compete with the present hybrid solutions a further reduction in costs achievable by a fully monolithic design is desirable.  The H35DEMO is a large electrode full reticle demonstrator chip produced in AMS \SI{350}{nm} HV-CMOS technology by the collaboration of Karlsruher Institut f\"ur Technologie (KIT), Institut de F\'isica d'Altes Energies (IFAE), University of Liverpool and University of Geneva. It includes two large monolithic pixel matrices which can be operated standalone.
One of these two matrices was characterised at beam test before and after irradiation with protons and neutrons.
Results demonstrated the feasibility of producing radiation hard large area fully monolithic pixel sensors in HV-CMOS technology. H35DEMO chips with a substrate resistivity of \SI{200}{\Omega cm} irradiated with neutrons showed a radiation hardness up to a fluence of \SI{1e15}{\neqcm{}} with a hit efficiency of about \SI{99}{\%} and a noise occupancy lower than \SI{e-6}{} hits in a LHC bunch crossing of \SI{25}{ns} at \SI{150}{V}.}

\begin{document}
\maketitle
\flushbottom

\section{Introduction}\label{sec:intro}
Monolithic Active Pixel Sensors (MAPS) produced with CMOS process are becoming of great interest for High Energy Physics (HEP) experiments.
In particular Depleted MAPS (DMAPS) first proposed in~\cite{peric1} give the possibility of having a fast monolithic detector with a large depleted region and reduced thickness for HEP applications. In DMAPS the signal is generated by charge drift instead of diffusion resulting in a faster charge collection with respect to normal MAPS and less prone to the effects of radiation damage.
This technology is being investigated for the upgrade of the ATLAS detector at the High Luminosity Large Hadron Collider (HL-LHC) with the aim of covering large areas in the outer pixel layers of the Inner Tracker (ITk) using cost-effective detectors~\cite{atlasTDR}. First prototypes of depleted active pixel sensors produced in H18 AMS~\cite{ams} CMOS technology and capacitive coupled to ATLAS FE-I4 chips have already demonstrated to resists large radiation fluences and thus be interesting candidates for the LHC experiments~\cite{h18irr}. The next step towards a final module for the ITk will be the development of a full scale monolithic prototype. The H35DEMO, was thus developed with the main purpose of demonstrating the feasibility of producing large area devices and investigate full size monolithic structures. For this first large area prototype the \SI{350}{nm} technology (H35) was chosen instead of the \SI{180}{nm} technology (H18) to contain production costs and investigate its use for less demanding radiation applications such as strip detectors.

\section{The H35DEMO demonstrator}\label{sec:h35demo}
\begin{figure*}[tbph!]
\centering
	\includegraphics[width=.5\textwidth]{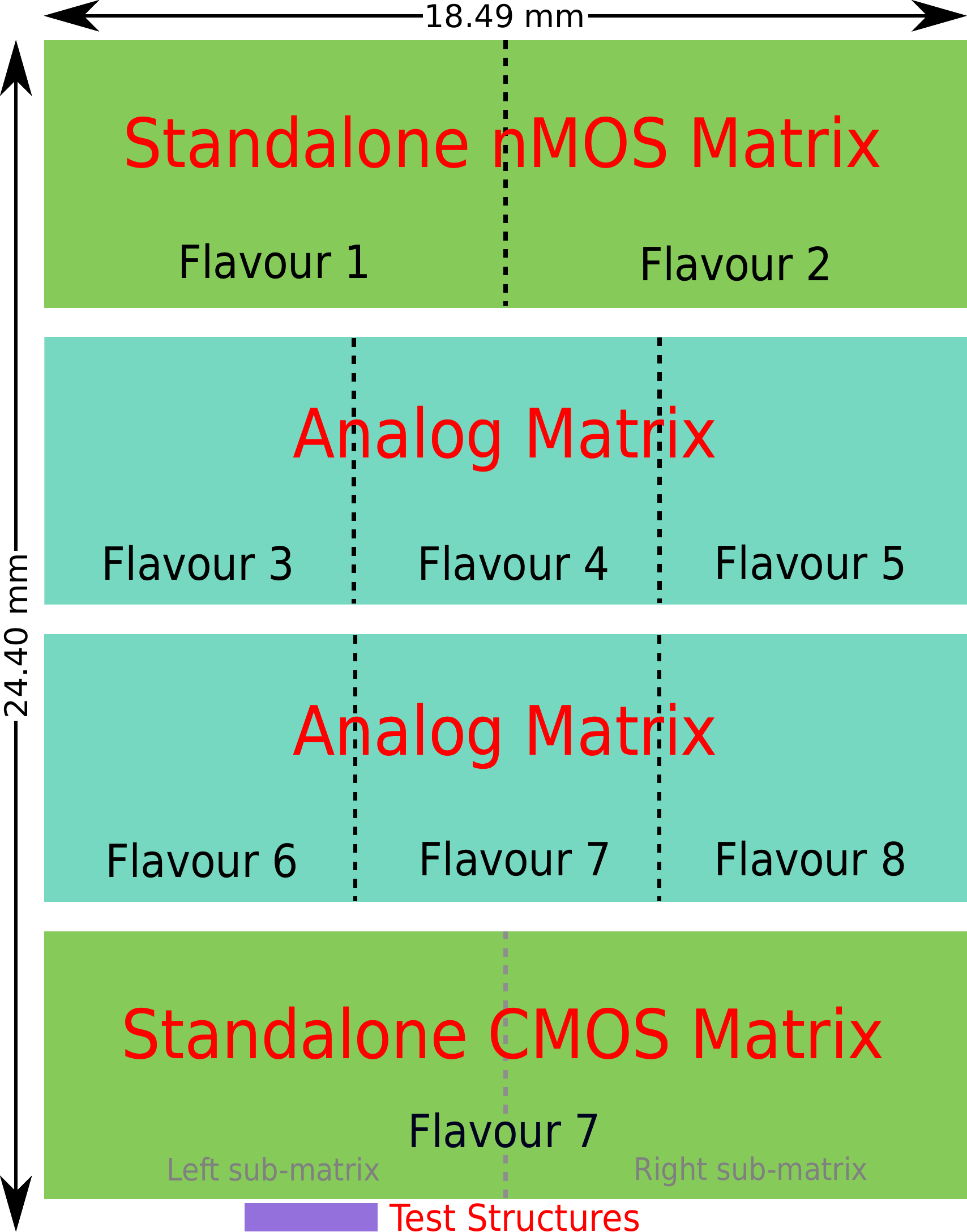}
\caption{Layout of the H35DEMO chip. The two large standalone matrices are represented in green, the two analog matrices in blue and the test structures in violet. The different flavours of pixels are highlighted: Flavours 1 (2) indicates pixels including discriminators without (with) time-walk compensation. Flavour 3 indicates pixels combining additional DP for high voltage bias and ELTs in the feedback circuitry. Flavour 4 (5) indicates pixels with no additional DP and ELTs (LTs) in the feedback circuitry. Flavour 6 (7) indicates high gain pixels with (without) additional DP for high voltage. Flavour 8 indicates low gain pixels without additional DP for high voltage.}
\label{fig:layout}
\end{figure*}

The H35DEMO~\cite{h35demo} is a large area demonstrator chip produced in AMS \SI{350}{nm} High Voltage CMOS (HV-CMOS) technology on four different substrate resistivities: \SIlist{20;80;200;1000}{\Omega cm}. The design was carried out by the collaboration of Karlsruher Institut f\"ur Technologie (KIT), Institut de F\'isica d'Altes Energies (IFAE) and University of Liverpool. The layout of the chip, shown in \fig{}~\ref{fig:layout}, consists of four independent pixel matrices: two analog matrices of \SI{23x300}{} pixels meant to be capacitive coupled to ATLAS FE-I4 readout ASICS~\cite{fei4}; and two monolithic matrices of \SI{16x300}{} pixels including digital electronics in the periphery to be operated standalone. Additional passive test structures without electronics are present at the edge of the chip. In particular a \SI{3x3}{} pixel matrix for Transient Current Technique (TCT) studies was extensively measured before and after irradiation, and results have been published in ref.~\cite{ecavalla}. 

The total dimensions of the chip are \SI{18.49x24.40}{mm}. In all four matrices the pixel size is \SI{50x250}{\um}, the same as in the FE-I4. The sensor part is a p-n junction obtained by a Deep N-Well over a p-type substrate. Analog electronics are embedded and shielded inside the same Deep N-Well also acting as collecting electrode. The chip is produced with a single side process on \SI{700}{\um} thick wafers with P-Well rings surrounding the N-Wells on the top surface where the negative High Voltage (HV) for reverse bias is applied.
The two analog matrices contain three different flavours of active pixels each combining the presence or the absence of additional bias rings between the Deep N-Wells, high or low gain, and Linear transistors (LT) or Enclosed Layout Transistors (ELTs).
An extensive description and characterisation of these matrices can be found in ref.~\cite{h35ccpd}. This paper will be dedicated to the characterisation of the monolithic part of the chip and in particular of the CMOS matrix. A detailed description of analog and digital architectures of the standalone matrices can be found in refs.~\cite{h35demo,raimonPoS}. In the following only the characteristics relevant to the presented measurements will be discussed.

\subsection{The monolithic CMOS matrix}
\begin{figure*}[tbph!]
\centering
\subfigure[Layout]{
	\includegraphics[width=.5\textwidth]{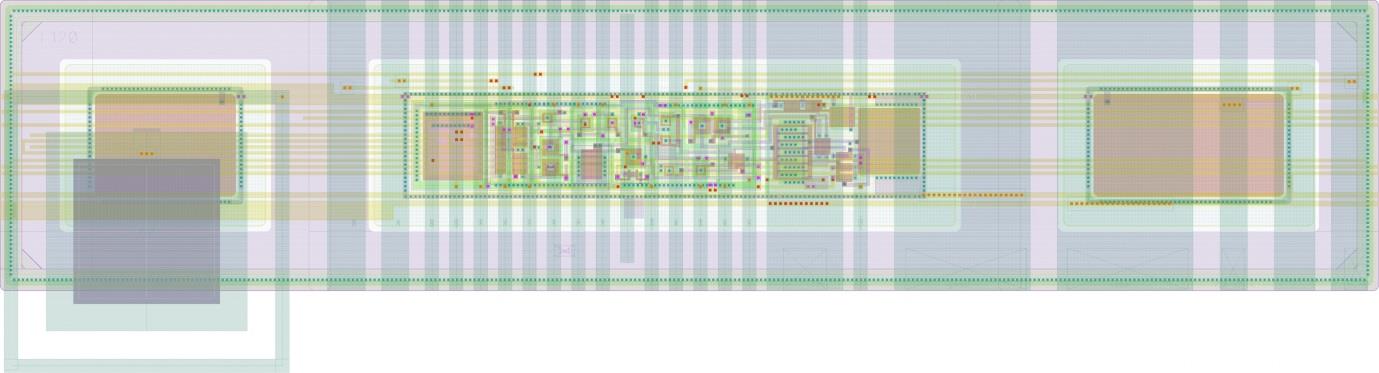}
	\label{fig:cmospixeltop}
}
\subfigure[Cross section]{
	\includegraphics[width=.55\textwidth]{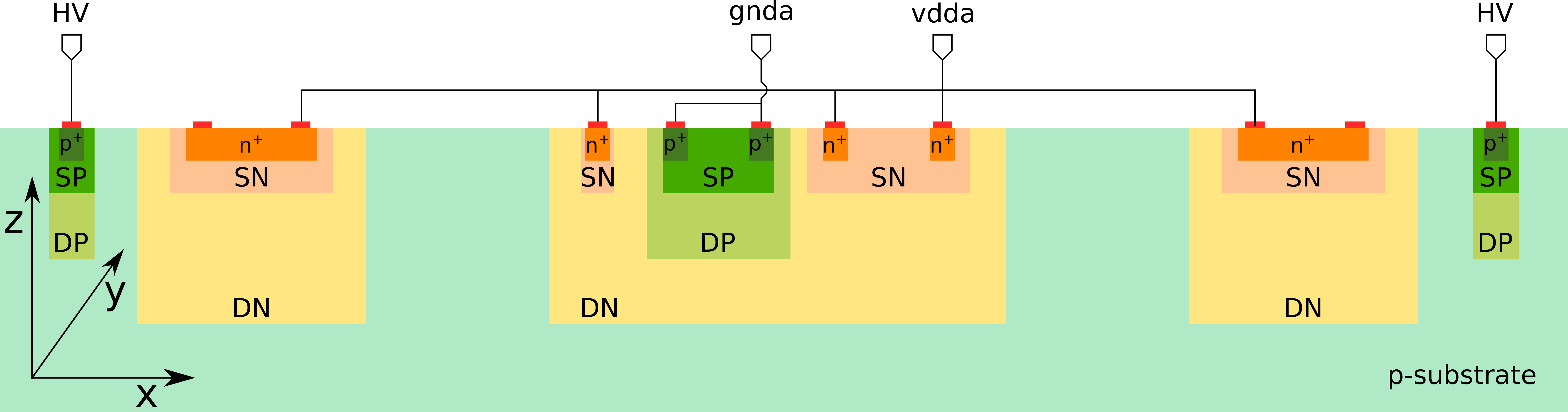}
	\label{fig:cmospixelside}
}
\caption{The pixel cell of the monolithic CMOS matrix. The pixel layout~\protect\subref{fig:cmospixeltop} and a sketch of the cross section~\protect\subref{fig:cmospixelside} are shown. In the latter DN and DP indicate Deep N-Wells and Deep P-Wells, respectively while SN and SP indicates shallow n and p-type implantations, respectively.}
\label{fig:cmospixel}
\end{figure*}

The pixels of the monolithic CMOS matrix combine high gain and LT without employing additional Deep P-Wells under the P-Well rings, as the pixels in the central columns of the second analog matrix (flavour 7). Layout and cross section of a pixel cell are shown in \fig{}~\ref{fig:cmospixel}. The Deep N-Well is divided into three parts to reduce the total capacitance of the pixel maintaining an uniform depletion of the bulk. The analog electronics are embedded in the central well and mainly include a Charge Sensitive Amplifier (CSA), a shaper and a second stage amplifier. 

\begin{figure*}[tbph!]
\centering
	\includegraphics[width=0.9\textwidth]{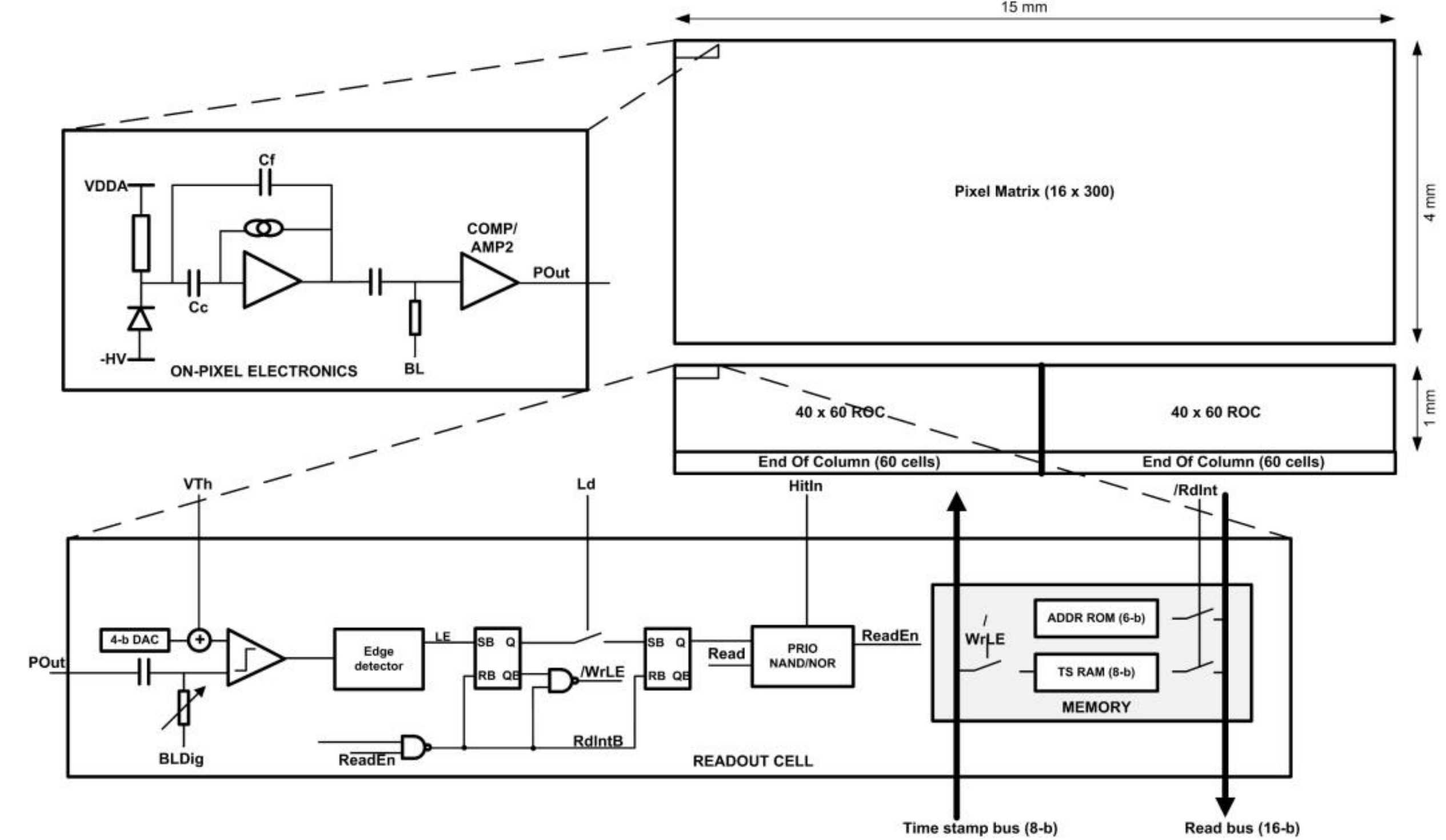}
\caption{Arrangement of the pixel CMOS matrix and its periphery with block diagrams of analog and digital front-end electronics. The analog front-end electronics are integrated on pixel while the digital electronics are placed in the periphery~\cite{raimonPoS}.}
\label{fig:block}
\end{figure*}

A sketch of the pixel and the periphery arrangement including block diagrams of the analog and digital electronics is shown in \fig{}~\ref{fig:block}. Each pixel is connected to a digital ReadOut Cell (ROC) placed in the periphery of the matrix where the response of the analog stage is processed. The ROCs are divided into two blocks arranged in matrices of 60 columns and 40 rows such that each ROC column is connected to two and a half columns of analog pixels. In the ROC a CMOS comparator is implemented to discriminate the signal generated by a crossing particle from the noise. When the signal crosses the comparator threshold, column, row and time-stamp relative to the threshold crossing time are stored. The right part of the matrix, i.e. columns from 150 to 299, includes an additional CMOS comparator. This can be set to a lower threshold to generate a second time-stamp which is stored only if the signal passed the threshold of the first comparator. The aim of this scheme is to improve the timing performance by reducing the size of the time-walk effect. The digital readout of the pixel matrix implements a column drain architecture with priority encoding where the information in the highest row is the first to be sent to the End Of Column (EOC). Data from the left most column of each half matrix is then serialised and sent out of the chip at the clock speed without zero suppression.

\subsection{The monolithic NMOS matrix}
The monolithic NMOS matrix integrates comparators made of NMOS transistors directly on pixel. The pixels in the left half of the matrix, consisting of columns from 0 to 149, employ standard NMOS comparators, while the pixels in the right half, columns 150 to 299, include more complex NMOS comparators to compensate for time walk by making the propagation time  independent of the amplitude of the input signal. Due to the space occupied by the additional electronics inside the pixel, the Deep N-Well is not separated as in the other matrices, leading to a larger capacitance. The digital front-end in the ROC is the same as in the left part of the CMOS matrix. In this case the CMOS comparators in the periphery allow to translate the low voltage output levels of the NMOS comparators to CMOS signals for the digital electronics.


\section{The readout system}\label{sec:readout}
\begin{figure*}[tbph!]
\centering
	\includegraphics[width=.5\textwidth]{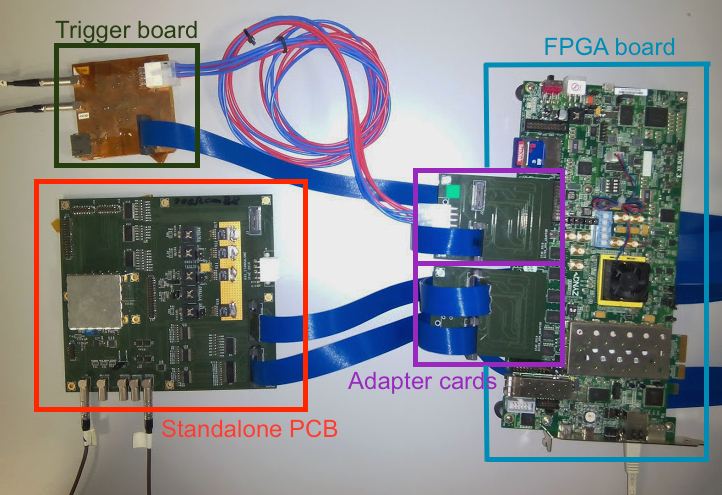}
\caption{The readout system developed at IFAE for the characterisation of the monolithic matrices of the H35DEMO chip.}
\label{fig:readout}
\end{figure*}

A Data Acquisition (DAQ) system dedicated to read out the monolithic matrices of the H35DEMO chip was developed at IFAE, Barcelona and is shown in \fig{}~\ref{fig:readout}. The hardware is based on the Xilinx ZC706 FPGA development board\footnote{Xilinx: All Programmable, \href{https://www.xilinx.com} [https://www.xilinx.com]} and a custom designed carrier board where the chip is attached and wire-bonded. This, so called \textit{Standalone PCB}, includes low voltage regulators to power the different matrices of the chip and Low Voltage Differential Signaling (LVDS) for communications. Through the Standalone PCB it is possible to apply the bias to the junction, provide an external test pulse for characterisation and tuning, and monitor the analog output of the CSA of pixel in the first column of both NMOS and CMOS matrices. Additional pins and connections on the board are implemented to monitor different signals of the chip for debugging purposes.

A first version of the Standalone PCB was produced which allowed to fully wire-bond and program the two monolithic matrices only. In this configuration the leakage current of the sensor was observed to be of the order of milliamps due to floating potentials in the pixels of the analog matrices. Even if it was possible to operate non-irradiated chips in these conditions, such large leakage current becomes particularly critical for irradiated devices due to the risk of thermal runaway. The problem was solved with the design of a second version of the Standalone PCB allowing for wire-bonding and programming of the analog matrices. All the presented results were obtained using this second version of the standalone PCB with the exception of the non-irradiated \SI{80}{\Omega cm} chip which was measured at beam tests on the first version of the standalone PCB. 

A \textit{Trigger Board} was developed to add external trigger capabilities to the system enabling integration with particle tracking telescopes for beam test measurements. This board consists of one LEMO connector to accept TTL trigger input signals and a second LEMO connector used to deliver a TTL busy signal. Both connectors are terminated on \SI{50}{\Omega}. A trigger-busy scheme is used to synchronise the H35DEMO data taking with an external DAQ reference system: After receiving a trigger pulse the H35DEMO DAQ system issues a busy signal to prevent the external reference system to further process and send triggers until all data of the H35DEMO is readout and processed. The same functions can be alternatively delegated to an RJ45 connector also present on this board. The DAQ is completed by a series of FMC adapter cards enabling connections between the FPGA board and the other components of the system.

A H35DEMO specific FPGA firmware was developed for the Xilinx board which allows to program the shift registers of all four pixel matrices independently and operate one of the two monolithic matrices at the time in \textit{continue acquisition mode} or in \textit{trigger acquisition mode}. In the trigger acquisition mode three parameters (trigger delay, event block and dead time) can be tuned to define the time window in which the output data from the chip is stored following a trigger signal. The firmware also implements zero suppression and supports Direct Memory Access (DMA). 

A software written in C++ and using a Graphical User Interface (GUI) written in QT~\footnote{Qt Group, \href{https://www.qt.io}{www.qt.io}} is used to steer and synchronise the operations of the FPGA, the external power supplies, and the pulse generator, for debugging, tuning and data taking. It communicates with a server running on the on-board operating system of the ZC706 via ethernet connection using a TCP/IP protocol. 

\begin{figure*}[tbph!]
\centering
\subfigure[Test injection scan]{
	\includegraphics[width=.47\textwidth]{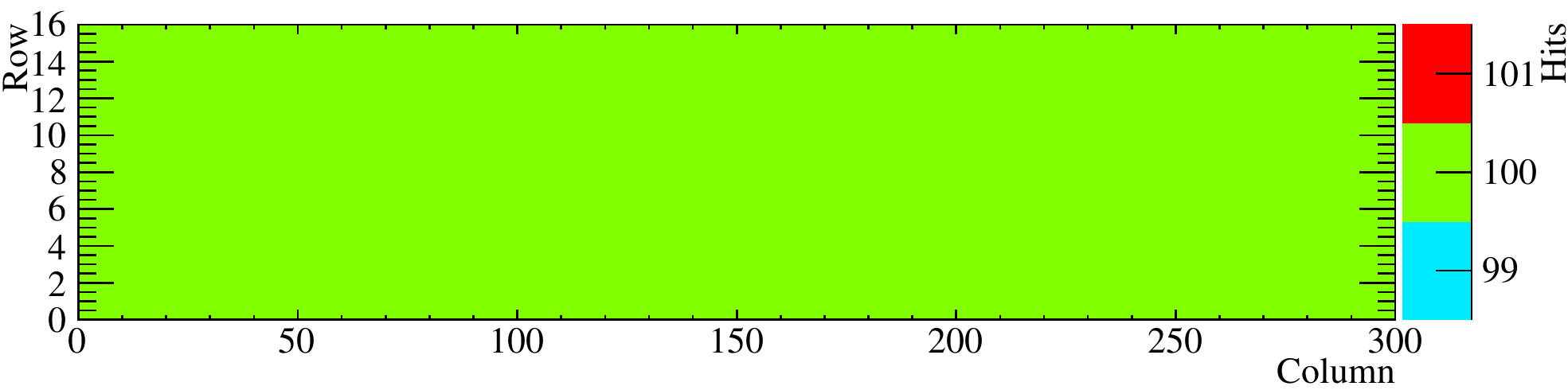}
	\label{fig:analogscan}
}
\subfigure[Source scan]{
	\includegraphics[width=.47\textwidth]{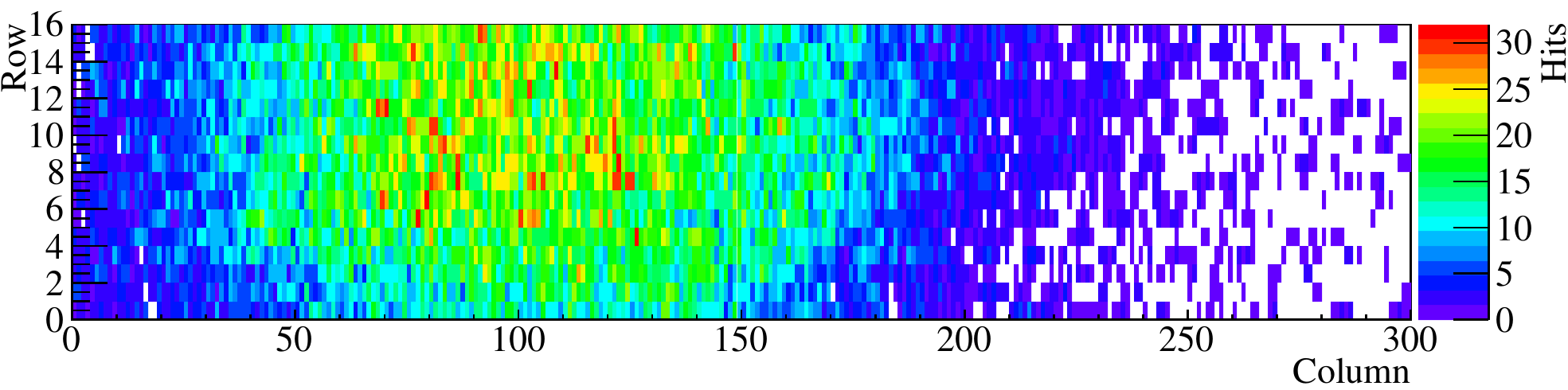}
	\label{fig:sourcescan}
}
\caption{Test of the functionalities of the CMOS monolithic matrix. In~\protect\subref{fig:analogscan} and~\protect\subref{fig:sourcescan} the response over the matrix surface to a test injection of 100 pulses per pixel and $^{90}$Sr beta electrons are shown, respectively.}
\label{fig:scans}
\end{figure*}

Examples of a \textit{test injection scan} and a \textit{source scan} obtained with this DAQ system for the CMOS monolithic matrix are shown in \fig{}~\ref{fig:scans}. In the test injection scan \SI{100} pulses per pixel are injected directly in the pre-amplifier while for the \textit{source scan} the signal was generated by a $^{90}Sr$ radioactive source of beta electrons placed on top of the matrix.

\section{Samples and irradiations}\label{sec:irrad}
H35DEMO chips with \SI{200}{\Omega cm} resistivity substrate were irradiated with neutron in the TRIGA Mark II research reactor of the Jo\v{z}ef Stefan Institute (JSI) in Ljubljana~\cite{jsiirr} up to a fluence of \SI{2e15}{\neqcm{}} and at the KIT Zyclotron~\cite{kitirr} with \SI{23}{MeV} protons up to a fluence of \SI{1e15}{\neqcm{}}, equivalent to a Total Ionisation Dose (TID) of about \SI{150}{Mrad}. For neutron irradiations at JSI there is also a low TID of about \SI{0.1}{Mrad} per \SI{e14}{\neqcm{}}, which is due to gamma emission~\cite{gammaJSI}. Two chips, one with a substrate resistivity of \SI{80}{\Omega cm} and one of \SI{20}{\Omega cm} were also irradiated with neutrons at JSI to \SI{1.5e15}{\neqcm{}}. After irradiation all samples were subject to an average annealing due to shipping and handling of about 1 week at room temperature before being measured. A list of the available modules and correspondent irradiation fluences can be found in \tab{}~\ref{tab:modlist}.

	\begin{table}[tbph!]
	\caption{Overview of the measured modules and their irradiations.}
	\begin{center}
		\begin{tabular}{|lcccc|}
			\hline
			Device         			& Resistivity  		  			&  Irradiation fluence 			& Irradiation 			& Beam test	    	 \\
			name				& $\mathrm{[\Omega cm]}$ 		& [\SI{e14}{\neqcm{}}] 			& facility				& facility \\
			\hline
			E3       					& 200					& 5					& JSI  				& FNAL\footnotemark  \\		
			E5       					& 200					& 5					& JSI  				& FNAL\footnotemark[\value{footnote}], SPS  \\	
			E7						& 200					& 10					& JSI				& SPS, DESY \\
			E10       					& 200					& 1					& KIT  				& DESY  \\		
			H7       					& 200					& 10					& KIT  				& DESY  \\	
			D5						& 200					& -						& -					& FNAL, SPS, DESY \\
			UG20-2					& 20						& -						& -	  				& FNAL  \\	
			UG80-1					& 80						& -						& -					& FNAL \\
			UG1k					& 1000					& - 						& -					& - \\
			UG20-1					& 20						& 15					& JSI  				& -  \\	
			UG80-2					& 80						& 15					& JSI				& -  \\
			D4       					& 200					& 15					& JSI  				& -  \\		
			D6       					& 200					& 15					& JSI  				& DESY  \\	
			D7						& 200					& 10					& JSI				& - \\	
			D9						& 200					& 20					& JSI				& DESY   \\
			\hline
		\end{tabular}
		\label{tab:modlist}
	\end{center}
	\end{table}

All irradiated devices were found to be programmable and working after irradiation, it was however not possible to measure all modules at beam tests due to the limited beam time and because of handling damage or assembly failures in some standalone PCBs.

\footnotetext{Results of irradiated modules from FNAL beam test are not reported in this paper, but can be found in ref~\cite{raimonPoS}.}

An analog injection test was performed to check the basic functionalities of the electronics after irradiation. This consists of injecting a certain number of large amplitude test pulses in each analog pixels and measure how many of these pulses are detected. In non-irradiated devices all the pulses are detected by all pixels, and the results of the scan is an uniform map of the pixel matrix as shown in \fig{}~\ref{fig:analogscan}. Since the injection is performed directly in the pre-amplifier, the effect of Non-Ionising Energy Loss (NIEL) in the silicon bulk for irradiated devices have negligible impact in this type of tests and thus only the behaviour of analog and digital electronics is investigated. 

\begin{figure*}[tbph!]
\centering
\subfigure[Pixel matrix representation]{
	\includegraphics[width=.7\textwidth]{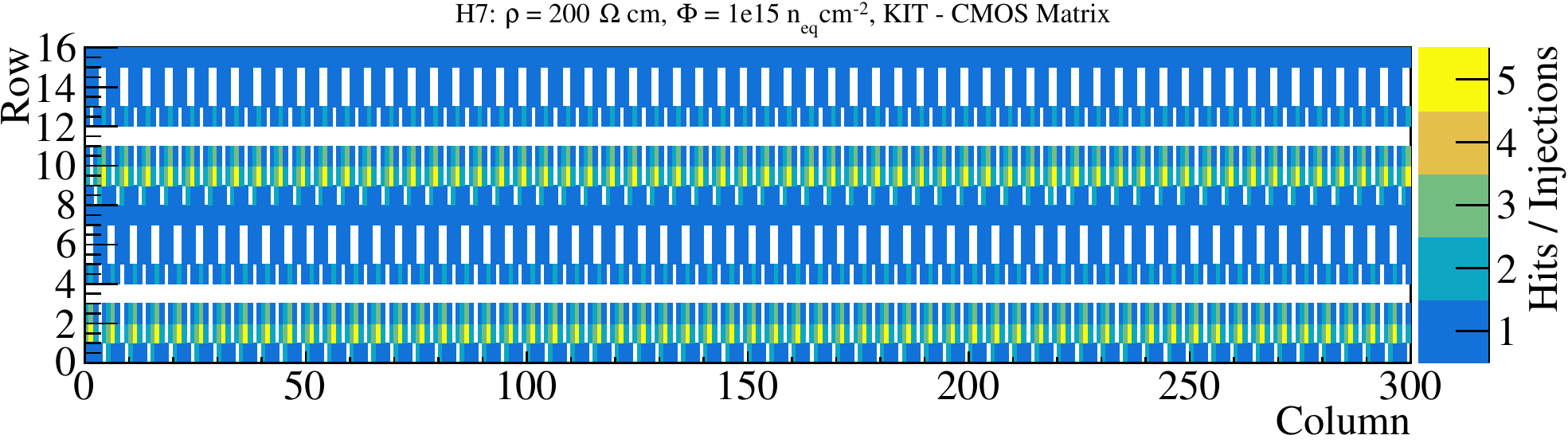}
	\label{fig:xtalk_ana}
}
\subfigure[ROC matrix representation]{
	\includegraphics[width=.45\textwidth]{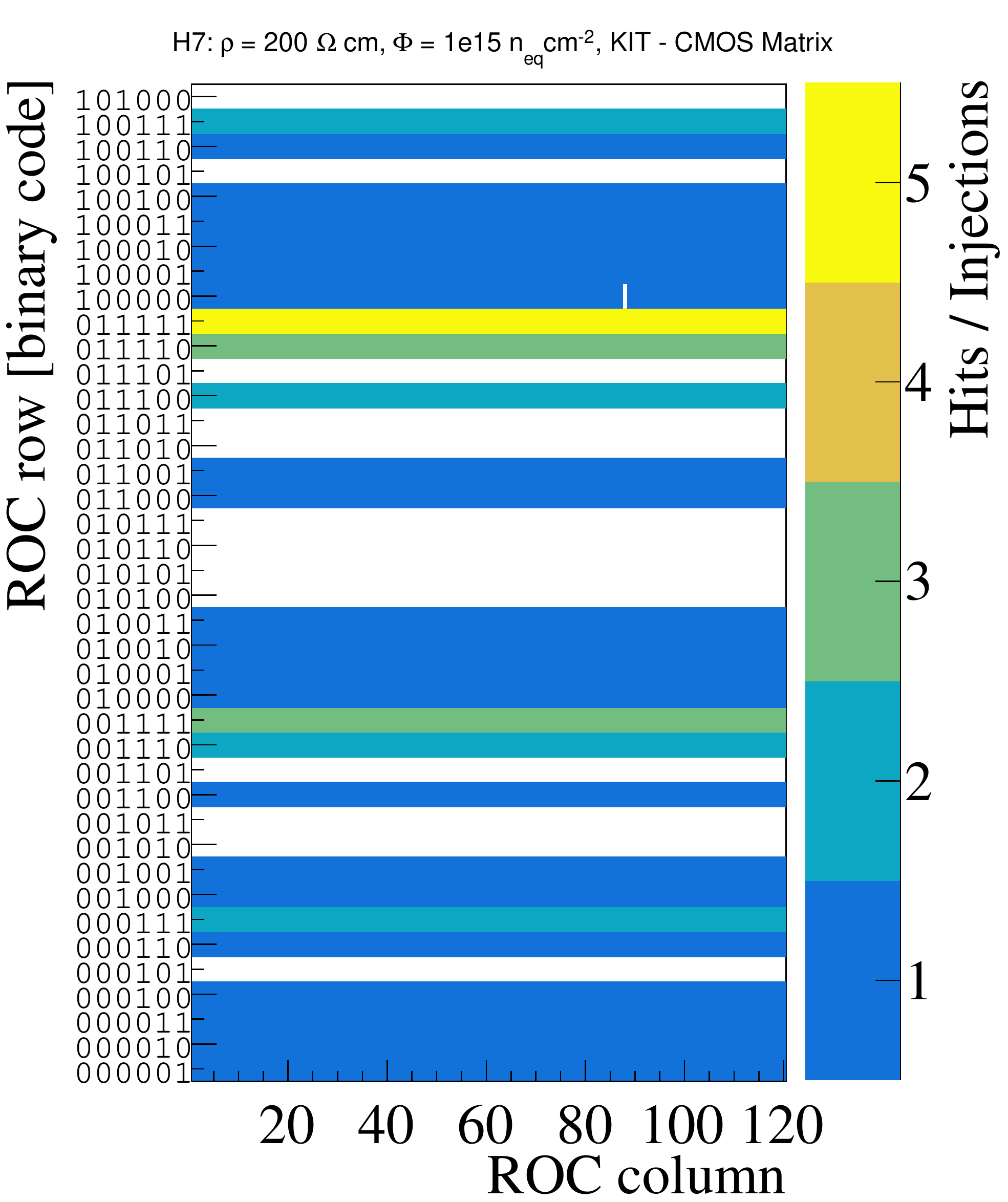}
	\label{fig:xtalk_dig}
}
\subfigure[Analog pixels to ROCs correspondence]{
	\includegraphics[width=.50\textwidth]{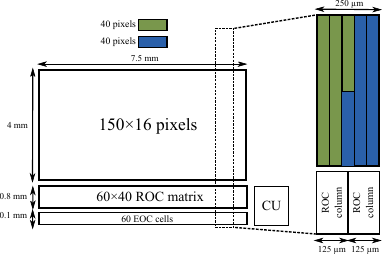}
	\label{fig:xtalk_sketch}
}
\caption{Analog scan of the CMOS matrix of a \SI{200}{\Omega cm} chip irradiated with protons to a fluence of \SI{1e15}{\neqcm{}}. In~\protect\subref{fig:xtalk_ana} the hits are shown in the pixel matrix representation, while in~\protect\subref{fig:xtalk_dig} the disposition of the pixels is shown as it is mapped to the ROC matrix. In both maps for each pixel the number of hits recorded is normalised to the number of injected pulses. The digital part of the chip is powered with \SI{3.3}{V}. The correspondence between each two and a half column of pixels in the analog part and the ROC columns in the digital part is sketched in~\protect\subref{fig:xtalk_sketch} for one sub-matrix.}
\label{fig:xtalk}
\end{figure*}

After neutron irradiation up to \SI{1e15}{\neqcm{}}, the whole matrix still responds to injection, while for larger neutron fluences and for proton irradiations some pixels are unresponsive and some others respond up to five times more than expected. In neutron irradiated chips this behaviour shows up especially when chips are operated at temperatures below \SI{0}{\celsius}. This effect is observed to be in general larger after proton irradiations probably due to the significantly higher TID with respect to neutron irradiations. The largest number of mis-behaving pixel is observed after proton irradiation to the fluence of \SI{1e15}{\neqcm{}}, the resulting analog injection test maps are shown in \fig{}~\ref{fig:xtalk}. As can be evinced from these maps, the effect cannot be associated with cross talk between pixels neither in the analog matrix (\fig{}~\ref{fig:xtalk_ana}) nor in the the ROC pixels of the digital periphery (\fig{}~\ref{fig:xtalk_dig}). In the latter it is otherwise evident that the pattern comes from a crosstalk between adjacent row address digital lines: When in the row address appears the pattern ''101'', the central bit, which is at 0, flips to 1 giving as output systematically the wrong row address.  Annealing was observed to increase the number of misbehaving pixels in all irradiated devices. Also neutron irradiated chips which did not present misbehaving pixels just after irradiation, began to show this crosstalk effect after accumulating few days at room temperature due to transportation and handling. This design problem was identified and corrected for the next generations of AMS prototypes for ATLAS. Anyhow, as a workaround to still perform measurements on the H35DEMO prototype it was possible to recover the uniformity of the chip response by increasing the digital voltage (VDDD) of the chip from \SI{3.3}{V} up to \SI{5}{V} depending on the irradiation levels and types. This prevents the cross talk, but at the same time increases the noise especially in the pixels close to the digital periphery.

\section{Current-Bias (I-V) characterisation}
\begin{figure*}[tbph!]
\centering
\subfigure[Before irradiation]{
	\includegraphics[width=.48\textwidth]{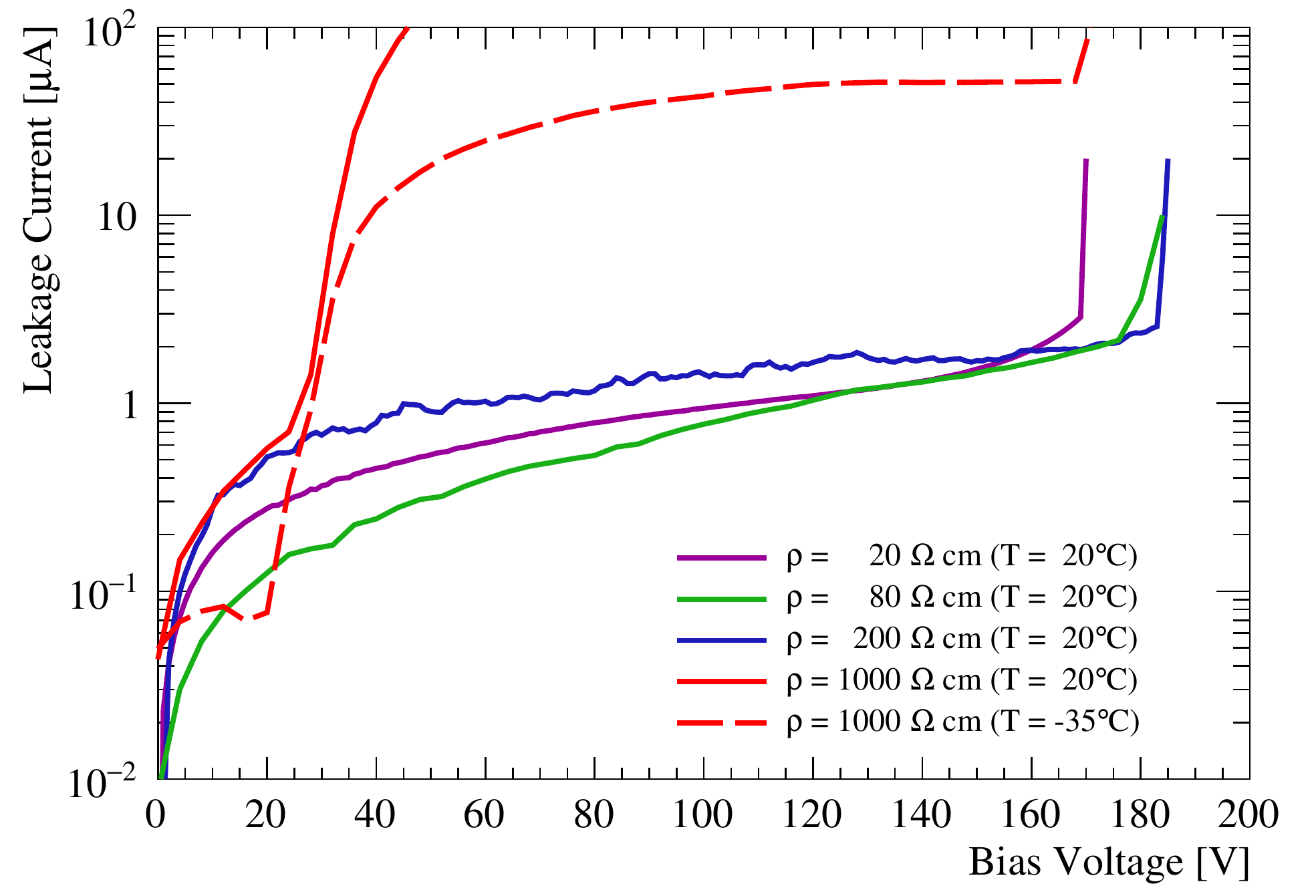}
	\label{fig:ivs_unirr}
}
\subfigure[After irradiation]{
	\includegraphics[width=.47\textwidth]{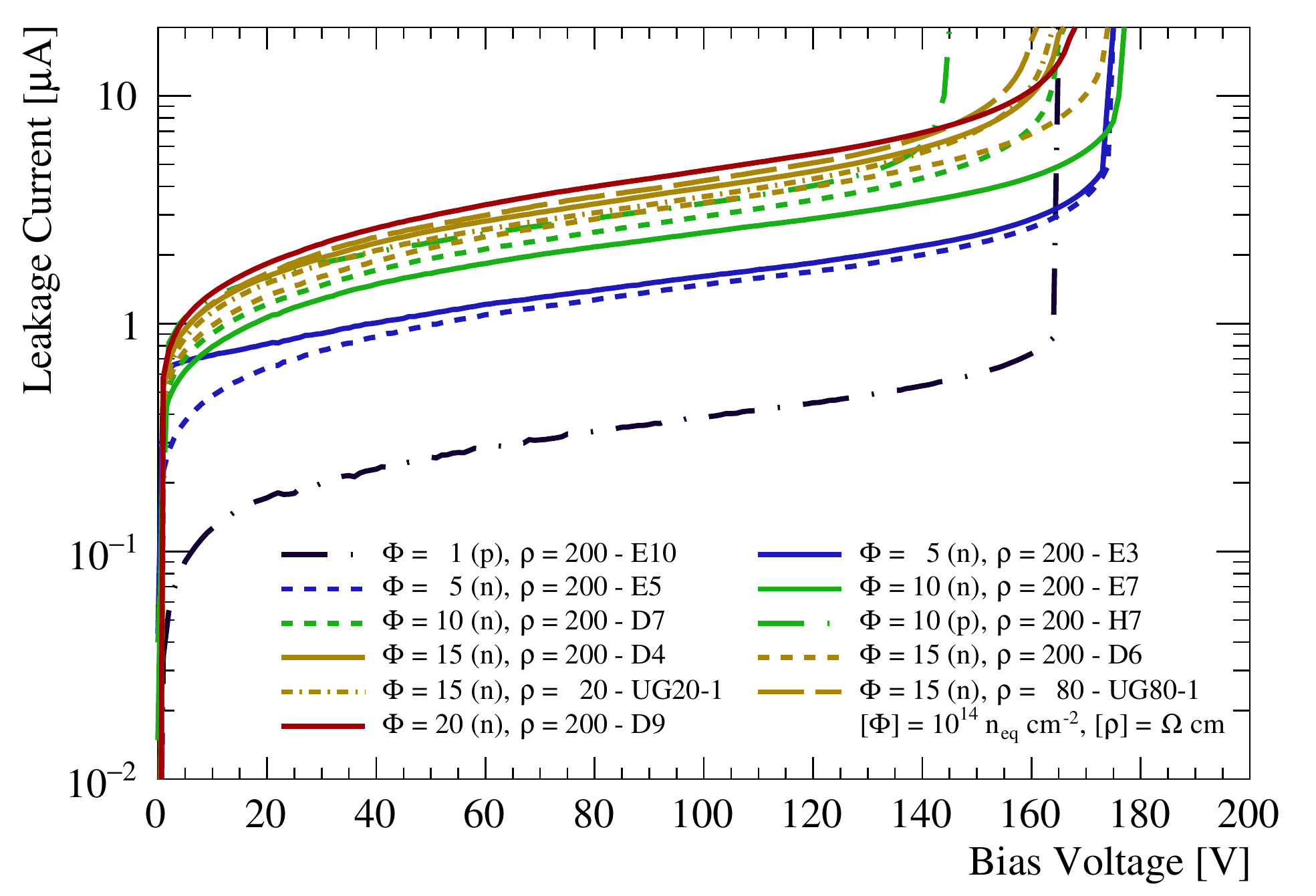}
	\label{fig:ivs_irr}
}
\caption{Current as a function of the bias voltage applied to the full device. Results of modules measured in a climate chamber before irradiation mostly at \SI{20}{\celsius} and after irradiation at \SI{-35}{\celsius} are shown in~\protect\subref{fig:ivs_unirr} and~\protect\subref{fig:ivs_irr}, respectively.}
\label{fig:ivs}
\end{figure*}

The leakage current of the p-n junction as function of applied bias (IV) was measured at stable temperature in a climate chamber. As shown in \fig{}~\ref{fig:ivs}, devices with a resistivity of \SIlist{20;80;200}{\Omega cm} experience the breakdown between \SIlist{165;185}{V} before irradiation and mostly above \SI{140} after irradiation. A steep increase of the current around \SI{30}{V} is instead observed in the \SI{1}{k\Omega cm} sample before irradiation. This was initially mistaken for a breakdown. Instead by measuring the module at \SI{-35}{\celsius}, thus reducing the leakage current below the instrument compliance to explore larger voltage biases, a second plateau is observed over \SI{60}{V}. After this step the real breakdown occurs at about \SI{165}{V} similarly to what observed for the other devices. This behaviour was also observed in \SI{1}{k\Omega cm} samples used for CCPD studies~\cite{h35ccpd}, and was explained by the Rise-And-Flatten (RAF) effect~\cite{RAF}. Due to the very high leakage current, it was not possible to operate and measure this device at beam tests. 

Irradiated devices show an increase of the leakage current with the fluence consistent with the linear expectation within the large uncertainties on the annealing times due to transport and handling.

\section{Threshold tuning and noise}\label{sec:tuning}
By setting dedicated internal Digital to Analog Converter (DAC) registers and applying external voltages through a DAC on the PCB, three main parameters can be tuned in the chip which have an effect on the threshold of the CMOS matrix. These are the feedback current of the in-pixel pre-amplifier (VNFBPix), the gain of the pre-amplifier (tuned by changing the voltage difference between nBLPix and ThPix) and the threshold of the discriminator in the periphery with respect to the baseline. The latter is set over the noise by changing the voltage difference between two global parameters (Th and nBL) which affect the whole matrix, and can be fine adjusted for each pixel with dedicated trim registers implemented in the chip. A \textit{trim global step} register (VPTrim) allows to define the amount of adjustment obtained by moving each trim register. The trim registers are disabled in the default configuration of the left sub-matrix, but not in the right one where a correction towards larger thresholds and proportional to the value of the VPTrim global register is always applied.
The tuning strategy for the presented measurements aimed at obtaining the lowest possible threshold on the left sub-matrix by changing the global parameters and subsequently rising the threshold on the right sub-matrix to minimise the noise level by increasing the VPTrim register. The fine tuning of the off-pixel trim registers of the CMOS matrix was not showing significant improvements in the threshold distribution and was therefore omitted.

\begin{figure*}[tbph!]
\centering
\subfigure[Off-pixel threshold]{
	\includegraphics[width=.47\textwidth]{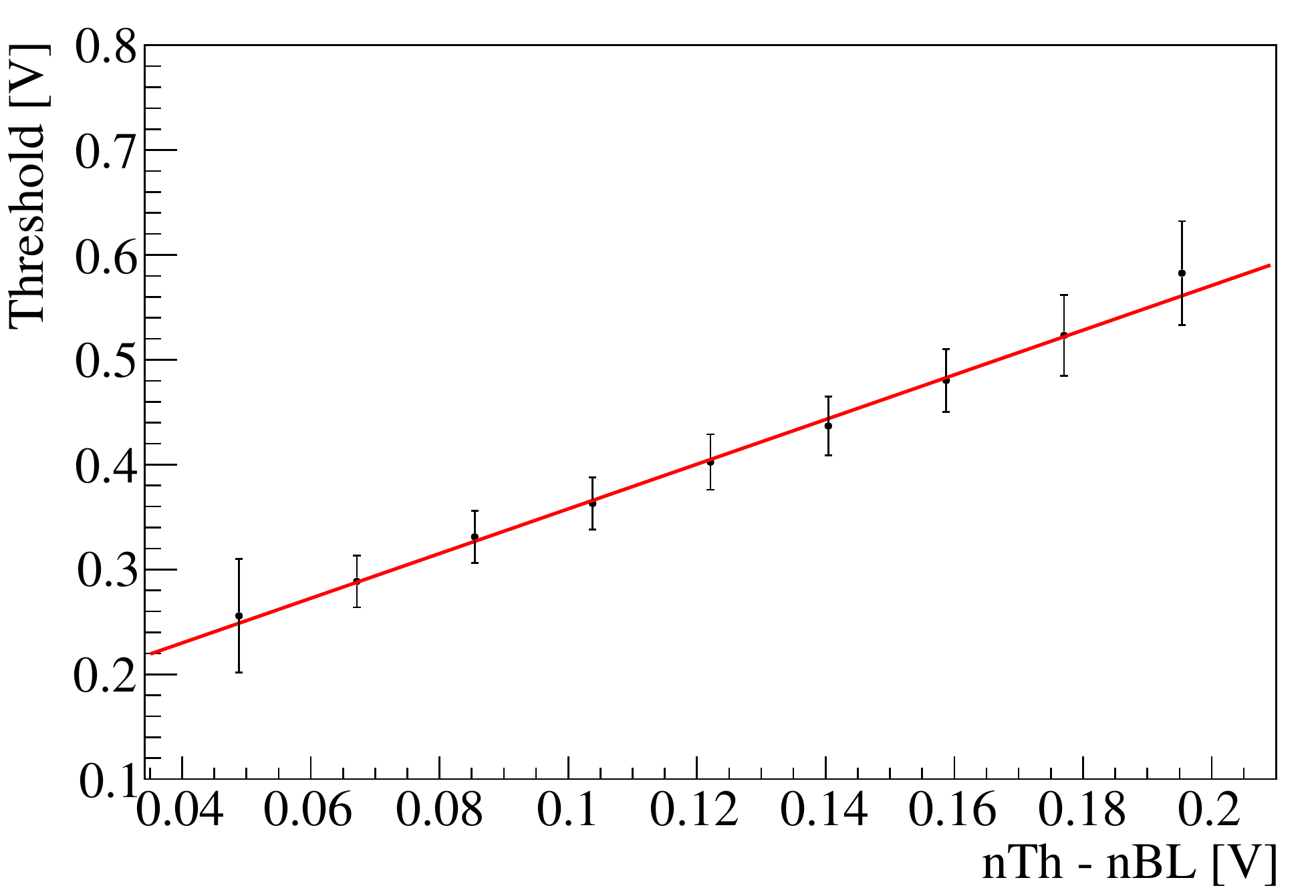}
	\label{fig:nth}
}
\subfigure[Gain]{
	\includegraphics[width=.48\textwidth]{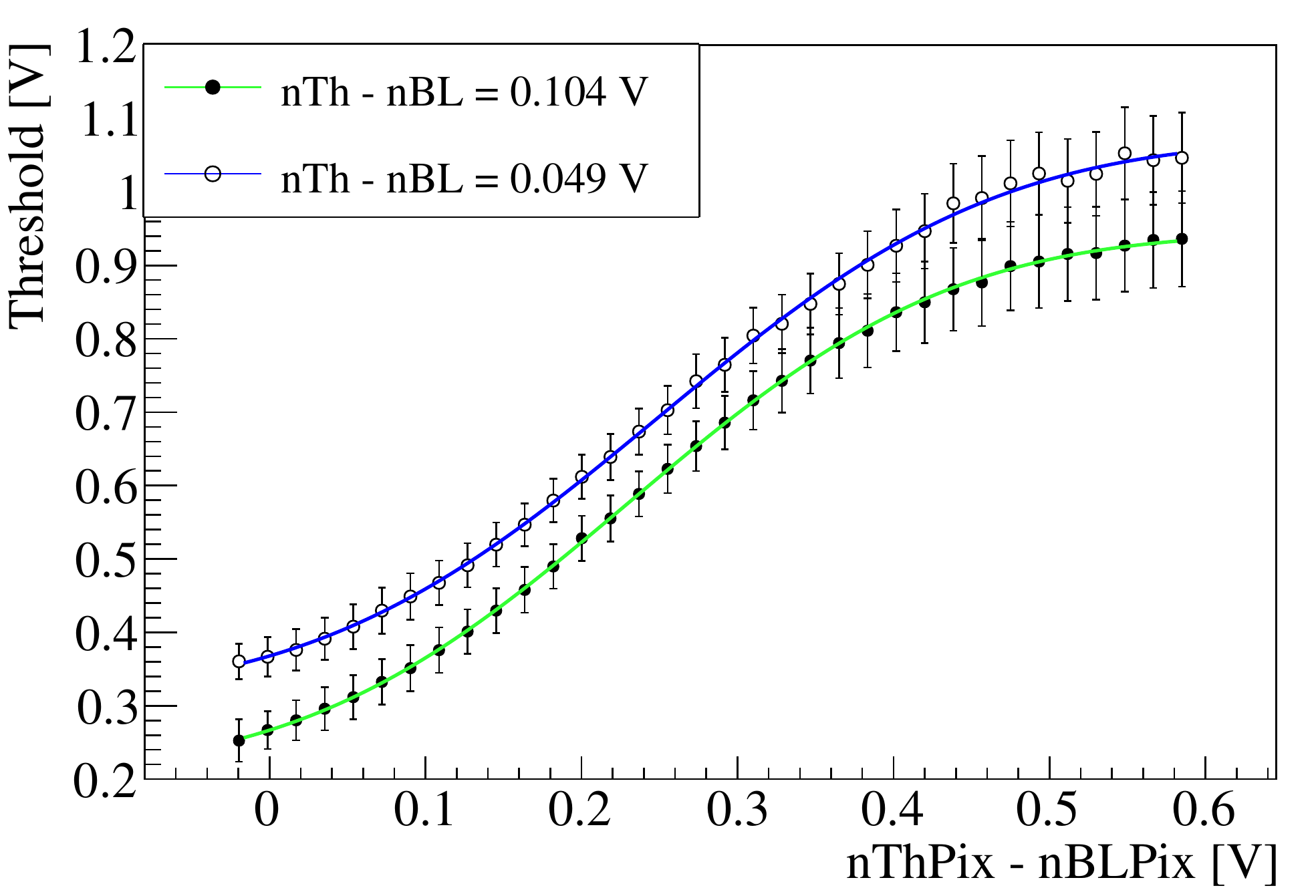}
	\label{fig:nthpix}
}
\caption{Mean threshold in the CMOS left sub-matrix as a function of the global parameter difference for the off-pixel threshold $\mathrm{nTh}-\mathrm{nBL}$~\protect\subref{fig:nth} and the gain $\mathrm{nThPix}-\mathrm{nBLPix}$~\protect\subref{fig:nthpix}. The threshold response to the gain is shown for two different settings of the off-pixel threshold. The uncertainties on the mean threshold value is the sigma of the Gaussian fit to the threshold distribution. The data points are fit with a first order polynomial and an error function for the off-pixel threshold and the gain, respectively.}
\label{fig:dacsbvsth}
\end{figure*}

The charge injection circuit included in each pixel allows to use an external test pulse to measure the \SI{50}{\%} occupancy point which defines the value of the threshold. 
The variation of the mean threshold distribution of the pixels in the CMOS left sub-matrix as a function of the global register settings is shown in \fig{}~\ref{fig:dacsbvsth}. When changing the off-pixel threshold settings a linear response is observed, while changing the global gain register the behaviour of the mean threshold distribution is well described in the measured range by an error function.

\begin{figure*}[tbph!]
\centering
	\includegraphics[width=.6\textwidth]{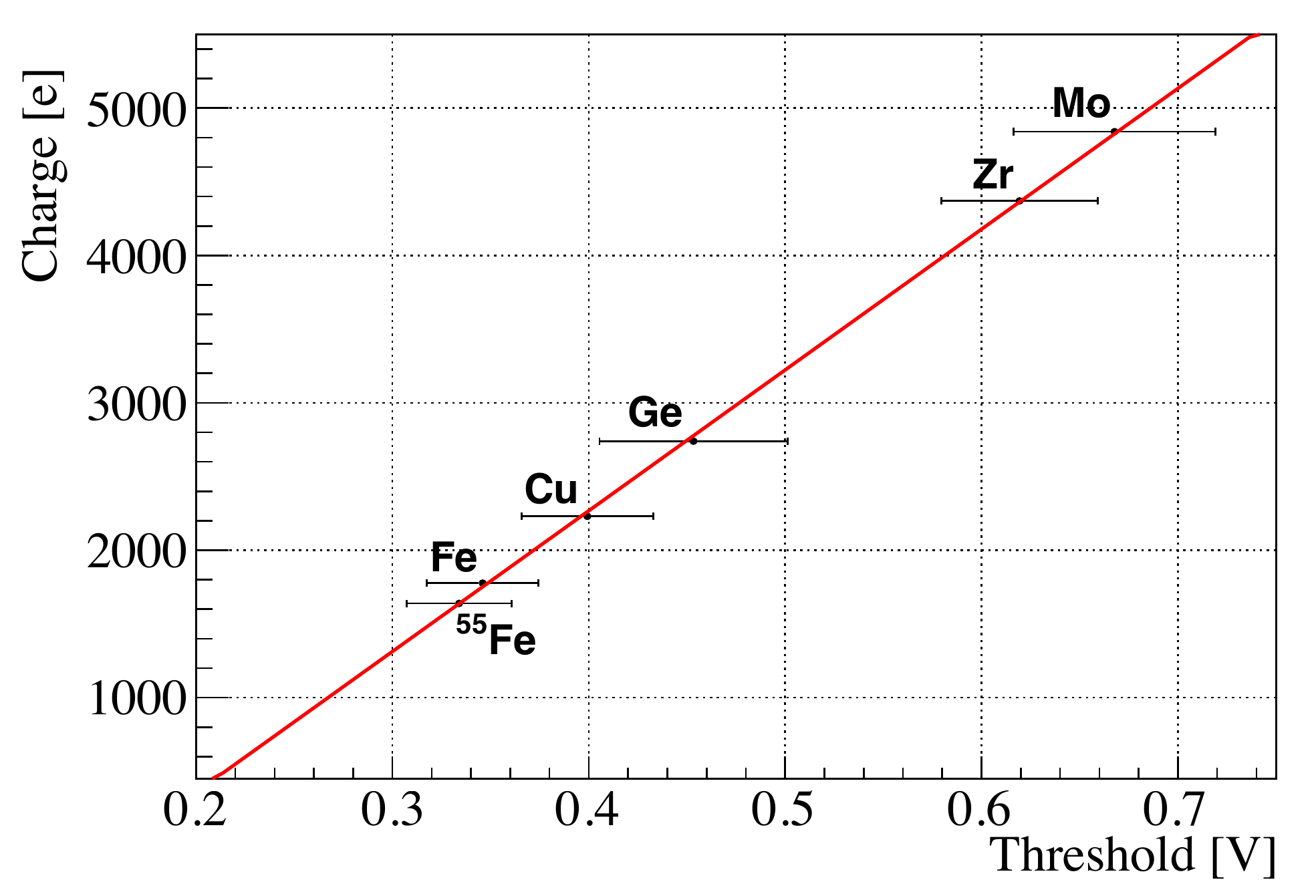}
	\label{fig:xray_calib}
\caption{Calibration of the mean threshold obtained by external voltage injection to charge in unit of number of electrons. The charge was generated in the sensor by the interaction in silicon of photons generated by x-ray fluorescence emission of several material: Iron (Fe), Copper (Cu), Germanium (Ge), Zirconium (Zr) and Molybdenum (Mo). An additional point was obtained using the gamma emission of an $^{55}\mathrm{Fe}$ radioactive source.}
\label{fig:xray_setup}
\end{figure*}

The correspondent equivalent injected charge was measured using an x-ray fluorescence setup at CERN with different target materials (Iron, Copper, Germanium, Zirconium, Molybdenum) and a $\mathrm{^{55}Fe}$ radioactive source emitting gammas. The integral of the photon spectrum is obtained by measuring the average response frequency (number of hits per second) of the pixels while scanning over the threshold range. This method relies on the assumption of full charge collection from photon interaction. The result of the measurements is shown in \fig{}~\ref{fig:xray_setup}. The calibration of the injected voltage to charge in unit of number of electrons extrapolated by a linear fit to the data is used in the following.

\begin{figure*}[tbph!]
\centering
\subfigure[Threshold]{
	\includegraphics[width=.47\textwidth]{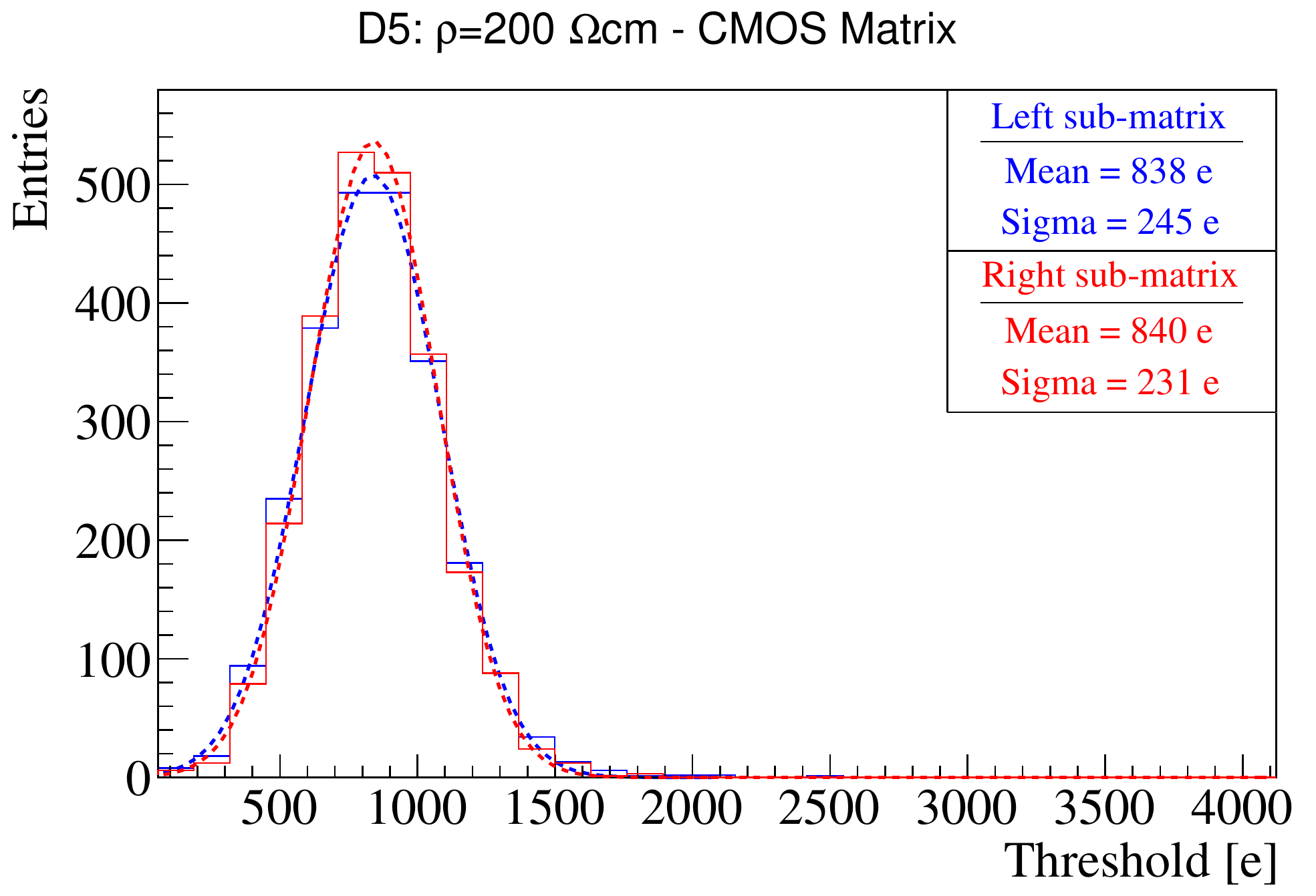}
	\label{fig:thr_unirr}
}
\subfigure[Noise]{
	\includegraphics[width=.47\textwidth]{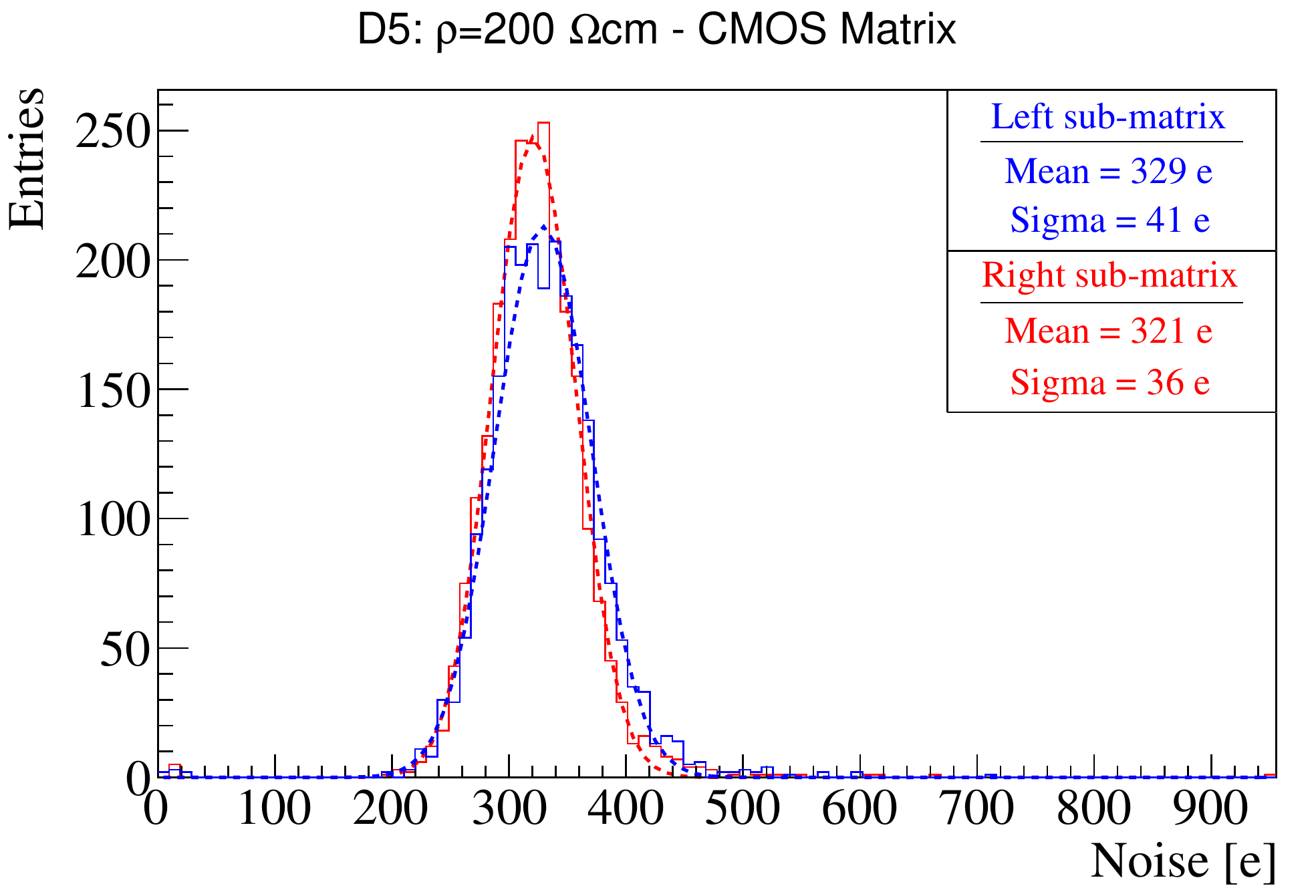}
	\label{fig:noise_unirr}
}
\caption{Threshold~\protect\subref{fig:thr_unirr} and noise~\protect\subref{fig:noise_unirr} distributions in the CMOS matrix of a non-irradiated detector with a bulk substrate of \SI{200}{\Omega cm}.}
\label{fig:thr_noise_unirr}
\end{figure*}

\begin{figure*}[tbph!]
\centering
\subfigure[Threshold]{
	\includegraphics[width=.47\textwidth]{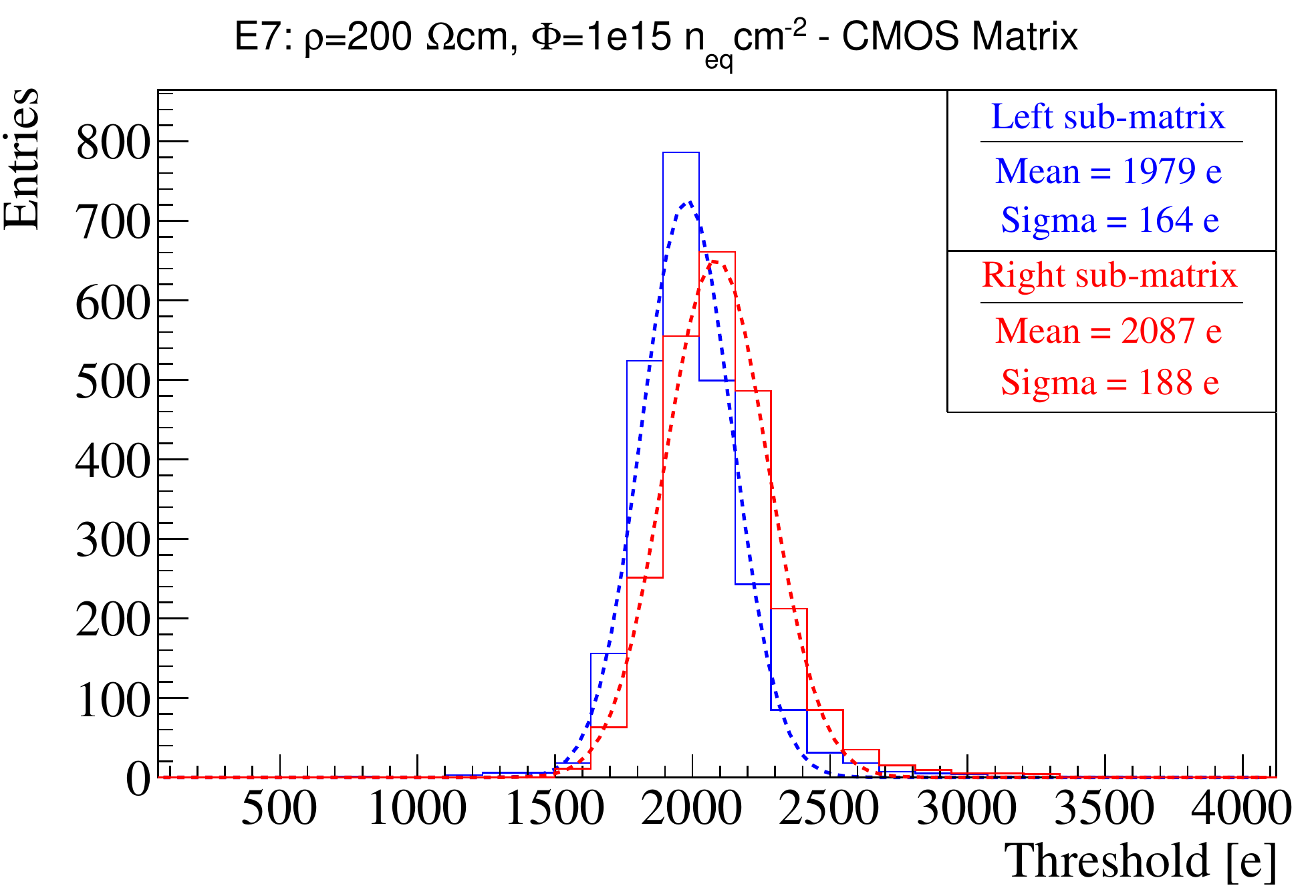}
	\label{fig:thr_1e15}
}
\subfigure[Noise]{
	\includegraphics[width=.47\textwidth]{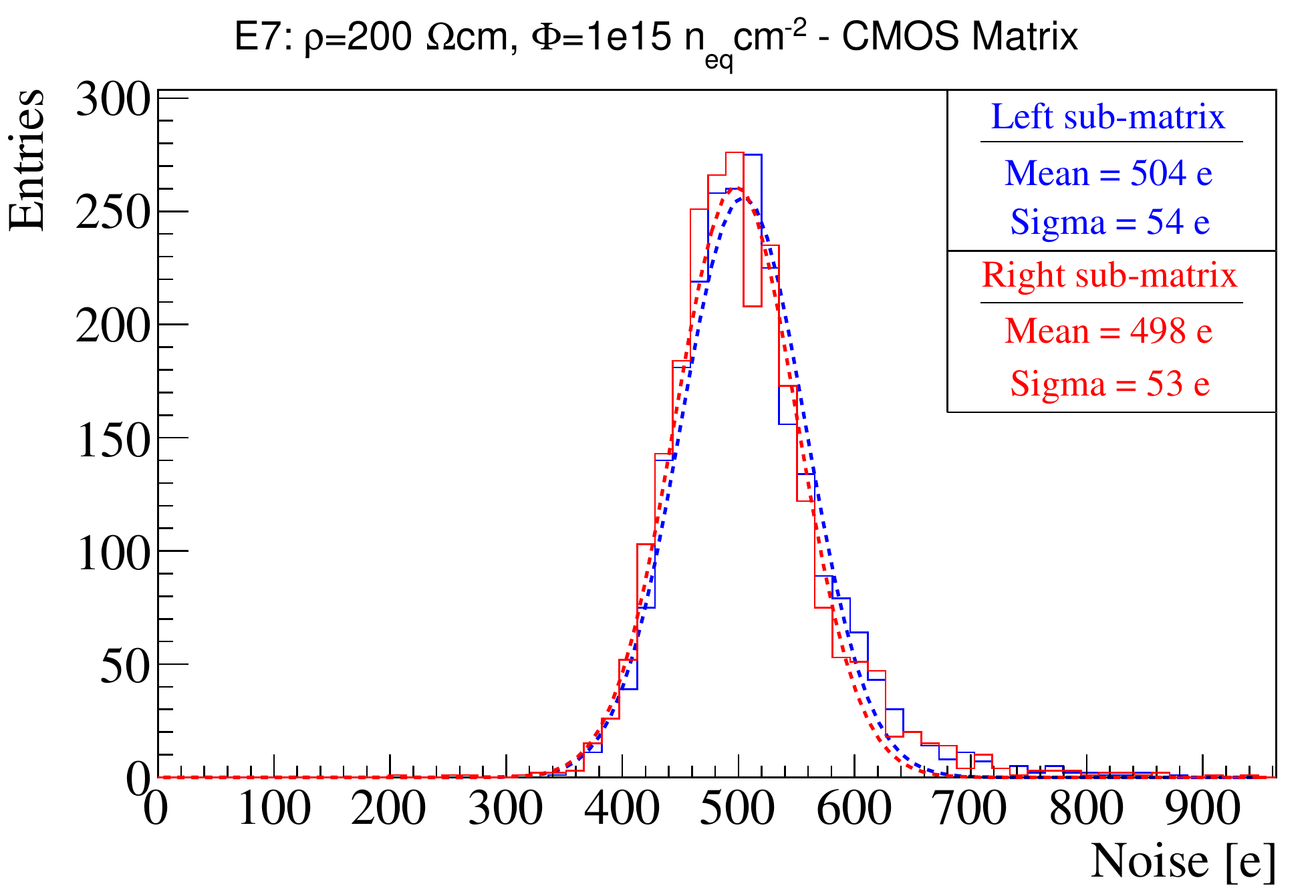}
	\label{fig:noise_1e15}
}
\caption{Threshold~\protect\subref{fig:thr_1e15} and noise~\protect\subref{fig:noise_1e15} distributions in the CMOS matrix of a detector with a bulk substrate of \SI{200}{\Omega cm} irradiated with reactor neutrons to a fluence of \SI{1e15}{\neqcm{}}.}
\label{fig:thr_noise_1e15}
\end{figure*}

An example of threshold tuning and correspondent noise for a \SI{200}{\Omega cm} chip before irradiation is shown in \fig{}~\ref{fig:thr_noise_unirr}. The module was kept at the constant temperature of \SI{20}{\celsius} inside a climate chamber.  A threshold of \SI{840}{e} with a sigma dispersion of about \SIrange{230}{250}{e} was obtained. Typical threshold noise values before irradiation are usually contained below \SI{500}{e} and in the case of the tuning presented the mean noise is about \SIrange{320}{330}{e} with a sigma of about \SI{40}{e}. 
         After irradiation the minimum achievable threshold is usually between \SIlist{1000;2200}{e}, depending on the irradiation dose, with a higher mean noise between \SIlist{500;600}{e} but mostly contained within \SI{700}{e}. An example is shown in \fig{}~\ref{fig:thr_noise_1e15} for a chip irradiated with neutrons to \SI{1e15}{\neqcm{}}. The lowest thresholds were obtained for the left sub-matrix only.

\section{Beam test campaigns: setups and analysis method}\label{sec:bt}
Beam test campaigns were carried out at the CERN Super Proton Synchrotron (SPS) in the H8 beam line with \SI{180}{GeV} pions, at the MTEST facility of the Fermilab National Accelerator Laboratory (FNAL) with \SI{120}{GeV} protons, and at Deutsches Elektronen-Synchrotron (DESY) with electrons and positrons between \SI{4}{GeV} and \SI{5}{GeV}.
At CERN and Fermilab the UniGe FE-I4 telescope~\cite{fei4telescope} was used as reference to measure the particle trajectories. It consists of six ATLAS FE-I4B planar pixel sensors from the IBL production read out with the RCE system~\cite{rce}. An AIDA EUDET-type telescope~\cite{eudet} made of six MIMOSA26 tracking planes and an FE-I4B planar pixel sensor as time reference was instead employed at DESY. With both telescope systems the trigger-busy signals described in \sect{}~\ref{sec:h35demo} were used to synchronise the data with the \textit{IFAE readout system}. In the case of the FE-I4 telescope the signals are handled by the RCE directly, while at DESY the AIDA TLU and the EUDAQ software are used to synchronise and merge the data of the EUDET telescope, the FE-I4B reference plane readout by a USBPix3 DAQ system~\cite{usbpix3}, and the IFAE readout system. 

In the FE-I4 telescope the trigger signal is given by the coincidence of hits in the first and last planes. As consequence of a trigger, hits in the telescope planes are integrated over a maximum time of \SI{400}{ns}, i.e. 16 LHC bunch crossings. In the IFAE readout system, instead, a time window of about \SIrange{7.5}{10}{\us} is selected to save the data coming from the H35DEMO chip. This large time window is necessary given the readout architecture of the chip in order to accommodate for the full readout of the EOCs of each half matrix containing the data related to the trigger event.

In the case of the EUDET telescope instead, particles are triggered by the coincidence of four scintillators overlapping over the area of a MIMOSA sensors, two are placed in front of the upstream arm and two behind the downstream arm. Since the integration time of the MIMOSA26 is much larger than the one necessary to read out the H35DEMO, additional tracks may be reconstructed which are not associated to the main particle triggered by the scintillators. To avoid an underestimation of the efficiency, an FE-I4 pixel module which has a triggered readout with a timing precision of \SI{25}{ns} is used as reference. For the analysis, only tracks pointing to a correspondent hit in the FE-I4 reference plane are selected.

For the reconstruction of reference tracks and extrapolation of the impact point on the detectors under test the Proteus software~\cite{proteus} was used for the FE-I4 telescope data, while for the EUDET telescope data the reconstruction and analysis frameworks EUTelescope and TBmonII were used~\cite{eutelescope,tbmon2}.

A report of the very first beam test campaign at CERN SPS in 2016 can be found in ref.~\cite{terzoINSTR17}. In this campaign the readout system was still a preliminary version showing stability problems and limitations that required a re-synchronisation of the events in the offline analysis. An upgrade of the readout system and a better understanding of the H35DEMO chip allowed to improve the integration with the FE-I4 telescope for better stability of the data taking. The results presented in this paper are obtained with this last version of the IFAE readout system.

\section{Beam test results}\label{sec:eff}
A full characterisation at different irradiation levels of chips with a substrate resistivity of \SI{200}{\Omega cm} was possible due to the large availability of these devices coming from the same wafer. Just few chips of other resistivities were instead available for these studies and thus were only measured and compared before irradiation.

\subsection{Operations and tunings}
Non-irradiated chips produced with three different resistivities, \SIlist{20;80;200}{\Omega cm}, were mounted on standalone PCBs and the CMOS monolithic matrix was measured in the beam at Fermilab. They were operated at room temperature with active heat dissipation using a fan blowing on the back of the device. A tuning of the CMOS matrix was performed aiming at the lowest threshold. After such tuning the left and right part of the matrix resulted in different threshold distributions and are thus considered separately in the analysis. The measured mean of the threshold distributions for the different chips was measured in both parts of the CMOS matrix and are summarised in \tab{}~\ref{tab:thr_unirr}. 

	\begin{table}[tbph!]
	\caption{Overview of the threshold distributions and digital voltage settings in the CMOS matrix for the chips measured at beam tests. Neutron and proton irradiations are indicated with \textit{n} and \textit{p}, respectively.}
	\begin{center}
		\begin{tabular}{|lccccc|}
			\hline
			Device          		&	Resistivity          			&  Irradiation fluence  & Mean threshold  			& Mean threshold 		& VDDD	    	 \\
			name 			&	$\mathrm{[\Omega cm]}$		&  [\SI{e14}{\neqcm{}}] (type)            			&  left $\mathrm{[e]}$	    		& right $\mathrm{[e]}	$	& $\mathrm{[V]}$\\
			\hline
			UG20-1    			&	20       					& 0						& 1350					& 1500  			& 3.3	 \\		
			UG80-2     		&	80       					& 0						& 1300					& 1700  			& 3.3	 \\	
			D5				&	200						& 0						&   800					& 1100			& 3.3 	\\ 
			E5       			&      200						& 5 (n)  					& 1700					& - 				& 3.3\\	
			E7				&      200 						&10 (n)					&  1700 					& - 				& 3.3\\
			E10       			&      200						& 1 (p)  					&  2100					& - 				& 3.9\\		
			H7       			&      200						& 10	(p)  					&  1700					& - 				& 4.5\\	
			D6       			&      200						& 15	(n)  					&  1700					& - 				& 3.9\\	
			D9				&      200 						& 20	(n)					&  2450					& - 				& 4.0\\	
			\hline		   
		\end{tabular}
		\label{tab:thr_unirr}
	\end{center}
	\end{table}

With the \SI{200}{\Omega cm} sample it was possible to reach thresholds as low as \SI{800}{e}. In the \SIlist{20;80}{\Omega cm} resistivity samples instead the tuning procedure resulted in a larger threshold of \SI{1300}{e} for the left part of the matrix and 1700 e for the right part of the matrix due to the noise that was found to be higher than for the \SI{200}{\Omega cm} chips. In the \SI{80}{\Omega cm} the larger noise may come from the large leakage current observed in the first version of the standalone PCB as explained in \sect{}~\ref{sec:readout}, while for the \SI{20}{\Omega cm} chips the higher noise may be connected with the substrate resistivity.

After irradiation the chips were operated at an environmental temperatures between \SIlist{-15;-25}{\celsius} using a chiller based cooling box at SPS, and between \SIlist{-35;-45}{\celsius} with dry ice at DESY. Due to the address crosstalk problem described in \sect{}~\ref{sec:irrad}, chips irradiated with neutron to fluences of more than \SI{1e15}{\neqcm{}} and proton irradiated chips were operated increasing VDDD as described in \tab{}~\ref{tab:thr_unirr}. 
Moreover, the threshold optimisation and the analysis was performed only for the left sub-matrix because of difficulties in obtaining an uniform tuning at low thresholds for irradiated chips in both left and right sub-matrices.

\subsection{Cluster properties}
\begin{figure*}[tbph!]
\centering
\subfigure[Before irradiation]{
	\includegraphics[width=.47\textwidth]{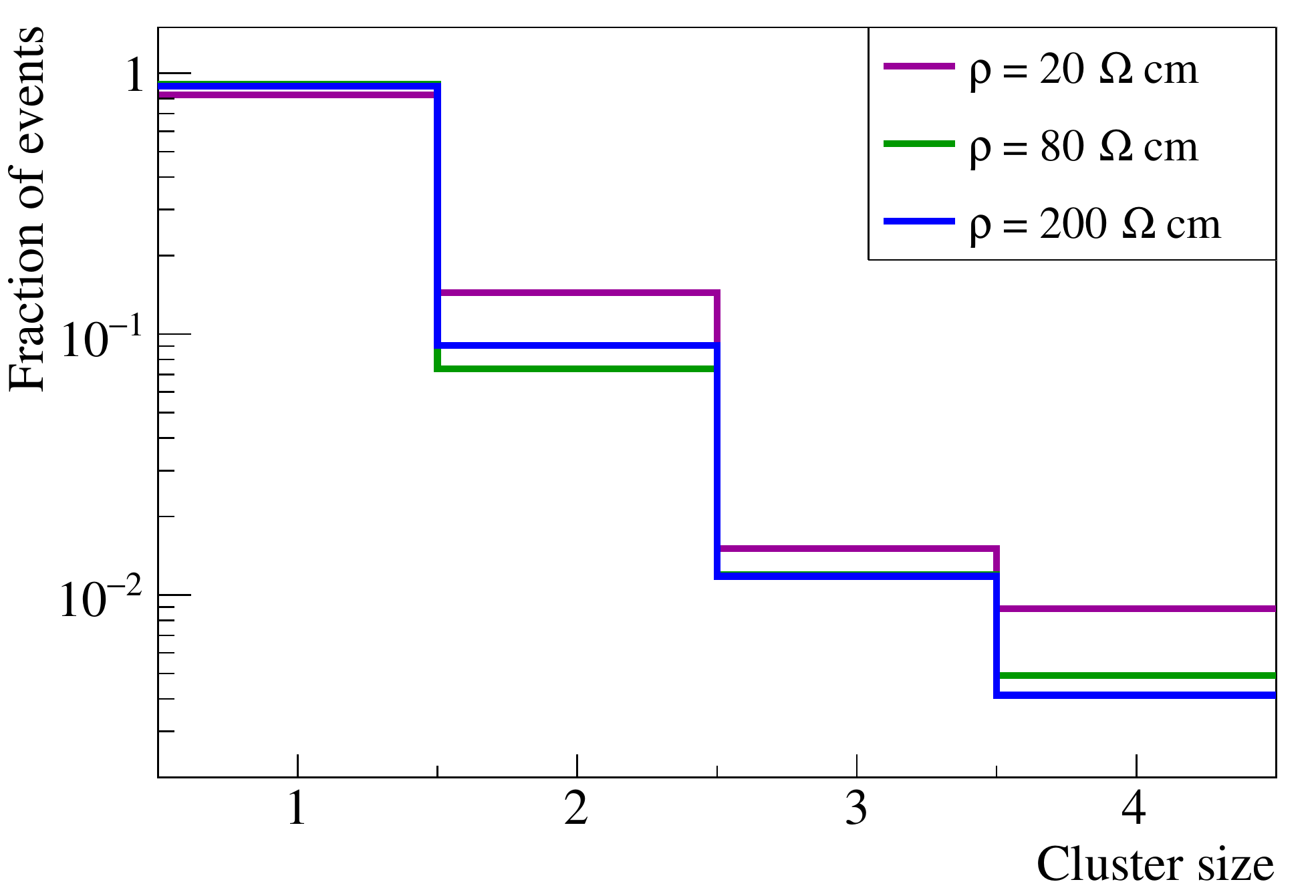}
	\label{fig:cs_unirr}
}
\subfigure[After irradiation]{
	\includegraphics[width=.47\textwidth]{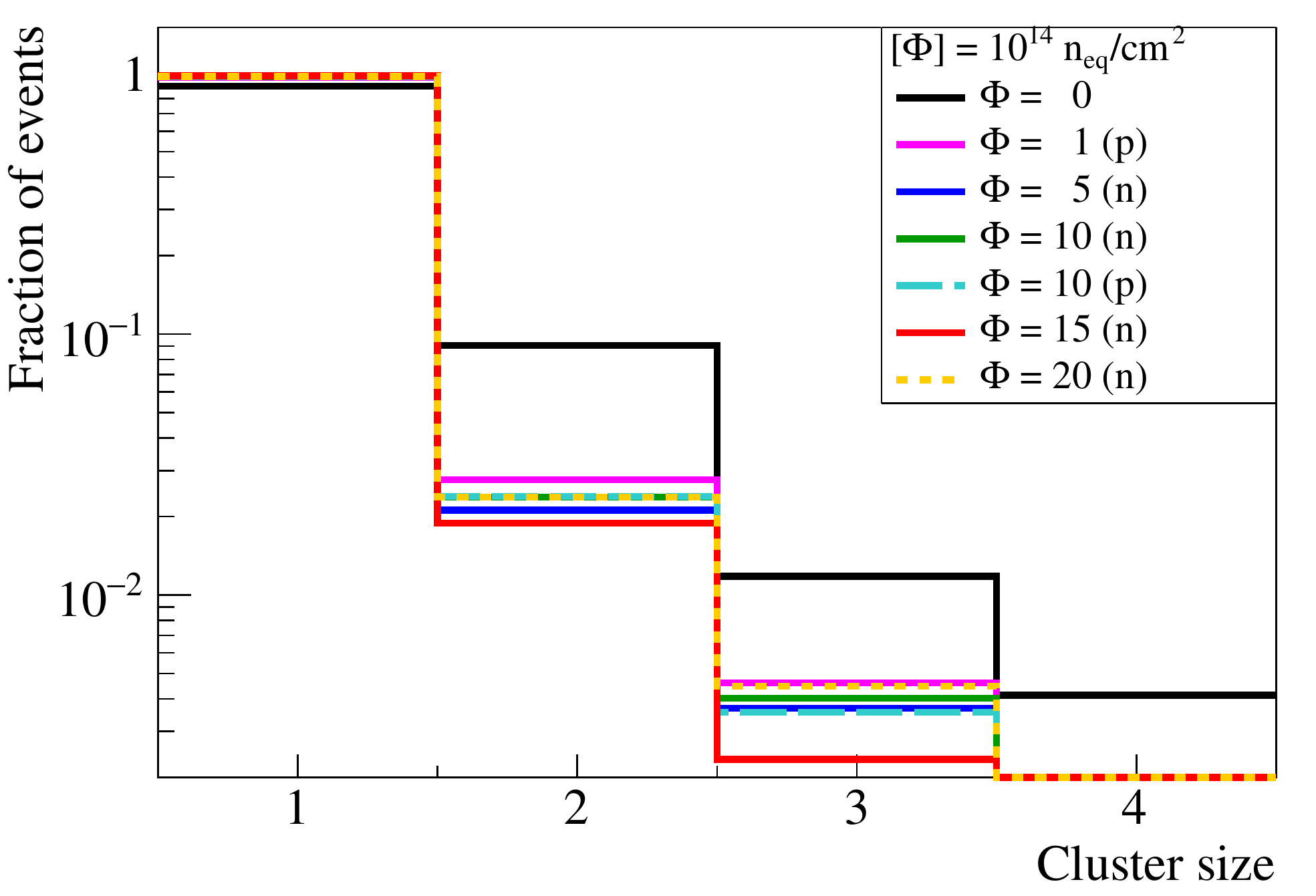}
	\label{fig:cs_irr}
}
\caption{Cluster size distribution for particles crossing perpendicularly to the detector surface. In~\protect\subref{fig:cs_unirr} the cluster size of non-irradiated samples of different resistivities is compared. In~\protect\subref{fig:cs_irr} results of \SI{200}{\Omega cm} samples after irradiation to different fluences are shown.}
\label{fig:cs}
\end{figure*}

The cluster size distribution measured at beam tests for modules before and after irradiation is shown in \fig{}~\ref{fig:cs}.
In these studies the beam was always impinging perpendicularly to the pixel surface, therefore a cluster size larger than one is expected only due to charge sharing out of lateral diffusion or eventually delta electrons. Before irradiation the fraction of cluster size two is about \SIrange{8}{9}{\%} in the \SIlist{80;200}{\Omega cm} samples, while in the \SI{20}{\Omega cm} a percent of clusters size two of about \SI{15}{\%} is observed. This difference is anyhow marginally significant given an uncertainty of up to 2 degrees in the alignment which leads to an uncertainty in the cluster fraction estimation of about \SI{5}{\%}.

After irradiation events with a cluster size larger than one are reduced by a factor 8-9 with respect to events before irradiations. This is expected due to charge trapping which reduces diffusion and the higher thresholds at which the chips were operated. The effect of the threshold variation on the cluster size was measured on the chip irradiated with neutrons to \SI{1e16}{\neqcm{}} and found to be less than \SI{1}{\%}.

\subsection{Efficiency before irradiation}

\begin{figure*}[tbph!]
\centering
\subfigure[Hit efficiency vs Bias Voltage]{
	\includegraphics[width=.53\textwidth]{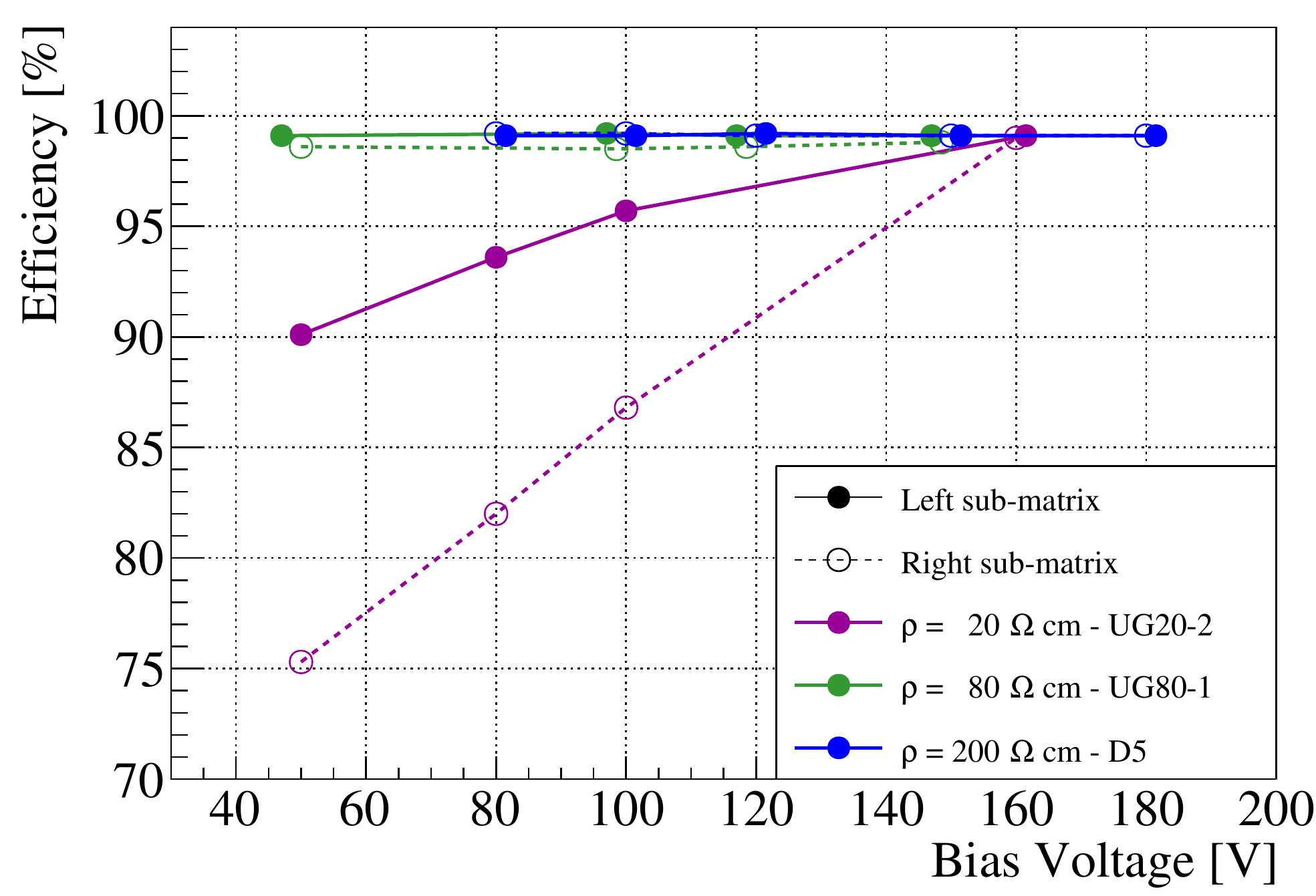}
	\label{fig:unirr_effvsV}
}
\subfigure[Hit efficiency maps at \SI{100}{V}]{
	\includegraphics[width=.42\textwidth]{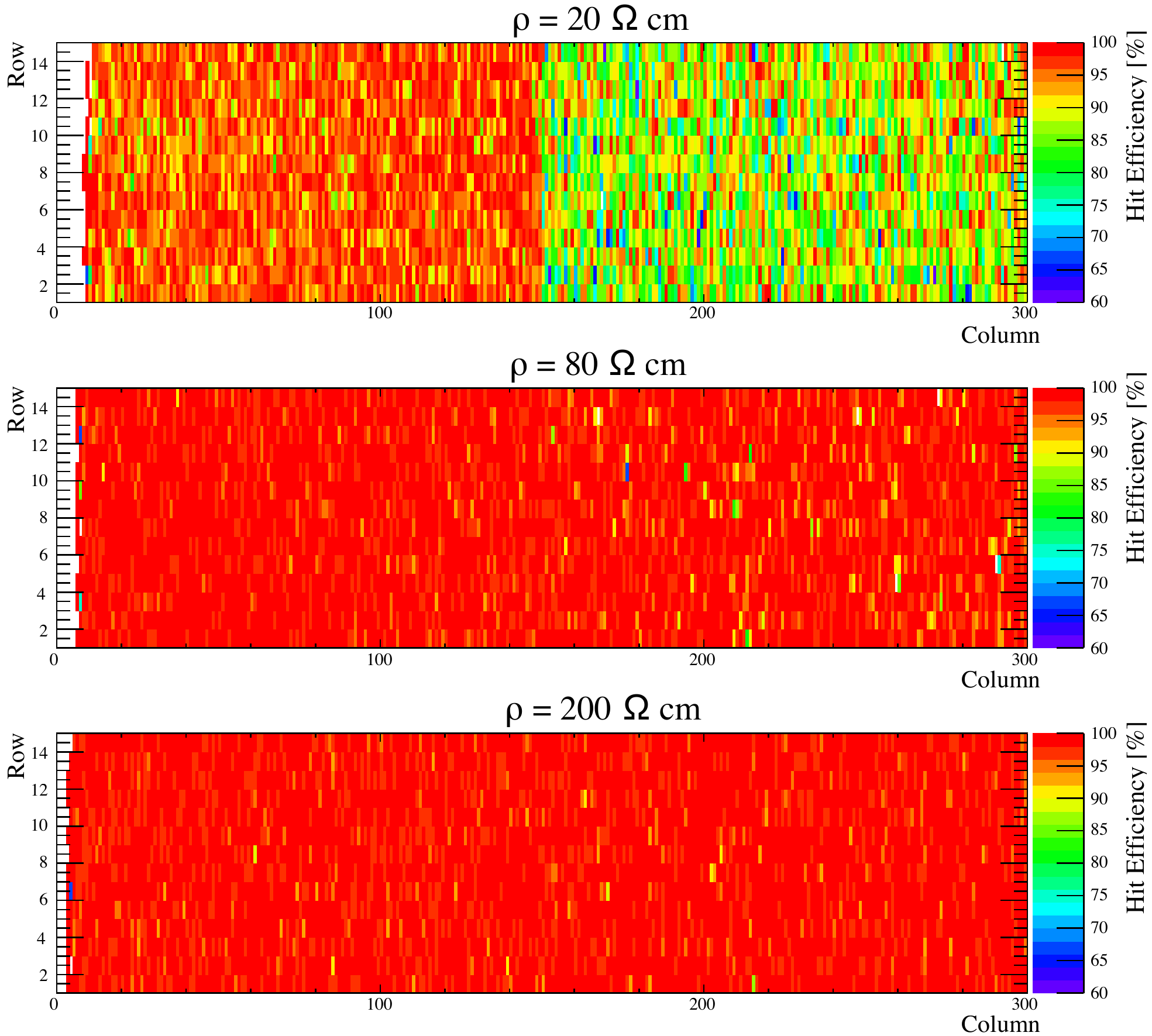}
	\label{fig:unirr_effmap}
}
\caption{Hit detection efficiency of the CMOS matrix before irradiation for different substrate resistivities. In~\protect\subref{fig:unirr_effvsV} the hit efficiency is shown as a function of the bias voltage. The performance of the left and right sub-matrices are shown separately due to difference in the threshold settings which became significant especially with low signals. Points are shifted of few Volts for better visibility. In~\protect\subref{fig:unirr_effmap} the hit efficiency distribution over the matrix surface at \SI{100}{V} is shown for three different resistivities (from top to bottom): \SIlist{20;80;200}{\Omega cm}. }
\label{fig:unirr_eff}
\end{figure*}

The hit efficiency measured as a function of the bias voltage is shown in \fig{}~\ref{fig:unirr_effvsV}. For resistivities of \SIlist{80;200}{\Omega cm} the hit efficiency is above \SI{99}{\%} already applying a bias voltage of \SI{50}{V}, while for the \SI{20}{\Omega cm} sample it was necessary to bias the junction with at least \SI{160}{V} to obtain the same results. This is explained with the lower depletion depth expected for such low resistivity~\cite{bern,ecavalla} which leads to a lower charge signal. Indeed, at lower voltages the hit efficiency decreases and a difference between the left and the right part of the CMOS matrix arises due to the slightly different threshold levels. This hit efficiency difference decreases as the bias voltage, and thus the signal, increases. This is the expected behaviour when the threshold is very close to the collected charge. 

Figure~\ref{fig:unirr_effmap} shows the hit efficiency distribution over the CMOS matrix when biased to \SI{100}{V}. In the \SI{20}{\Omega cm} sample and faintly also in the \SI{80}{\Omega cm} sample, it is possible to appreciate the different efficiency between left and the right sub-matrices which is uniform within each sub-matrix.
In the case of \SI{200}{\Omega cm}, instead, the hit efficiency is uniform over the whole pixel area.

\subsection{Efficiency after irradiation}
\begin{figure*}[tbph!]
\centering
\subfigure[Hit efficiency vs Bias Voltage]{
	\includegraphics[width=.47\textwidth]{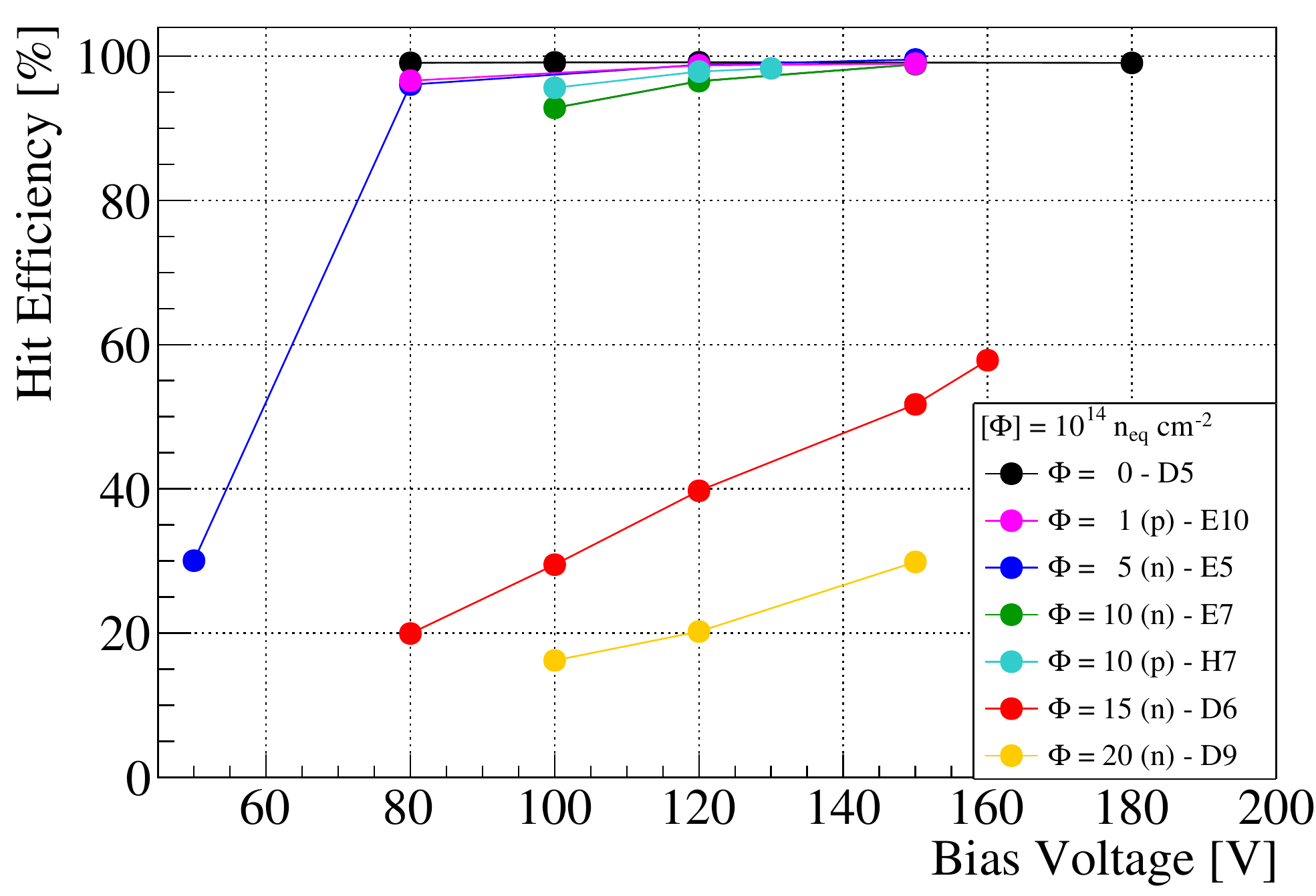}
	\label{fig:eff_irr_all}
}
\subfigure[Hit efficiency vs Bias Voltage, zoom]{
	\includegraphics[width=.47\textwidth]{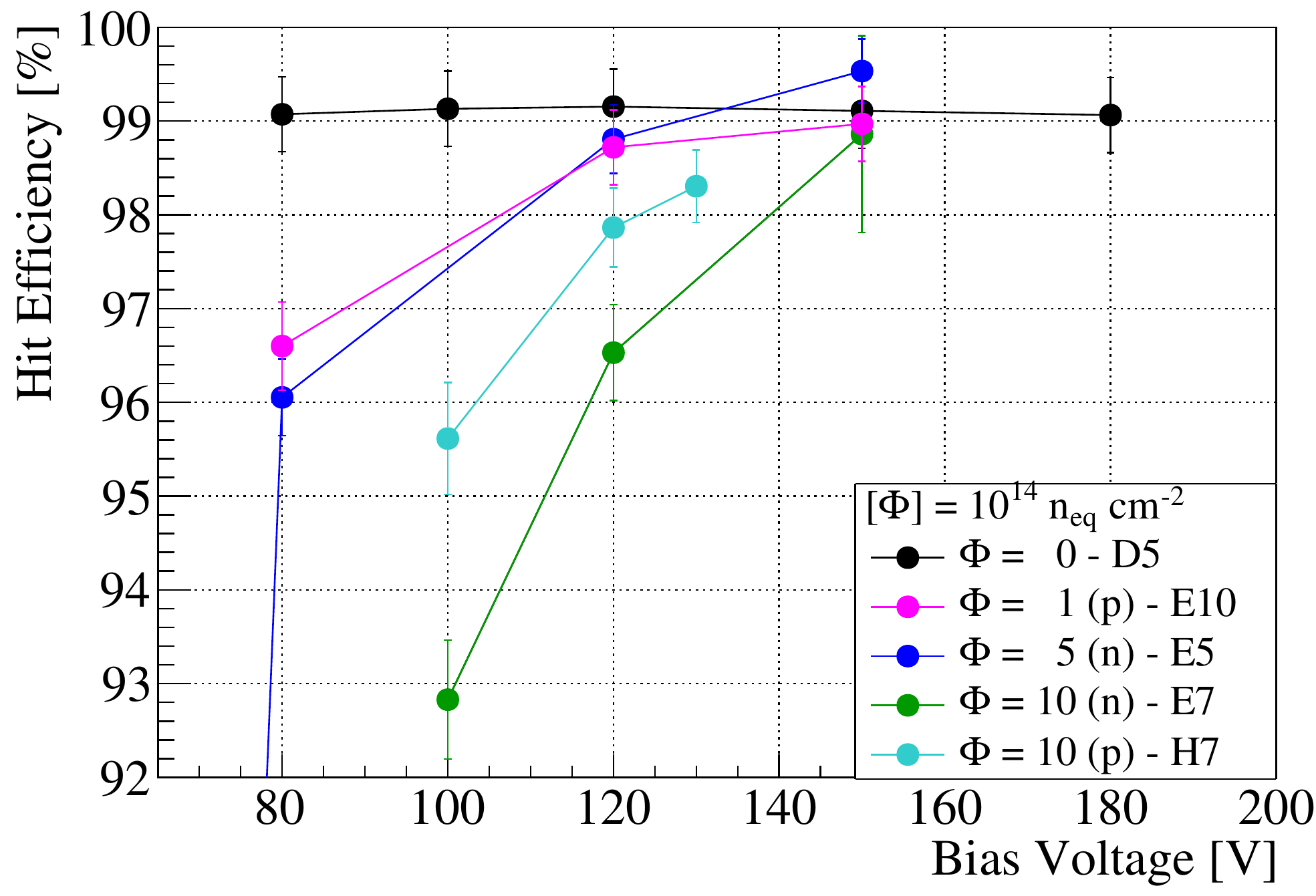}
	\label{fig:eff_irr_zoom}
}
\caption{Hit detection efficiency as a function of the bias voltage of the CMOS matrix for samples irradiated to different fluences. In~\protect\subref{fig:eff_irr_all} the full hit efficiency scale is shown, while \protect\subref{fig:eff_irr_zoom} shows a zoom over the high efficiency region. Neutron irradiation is indicated by (n) and proton irradiation by (p) in the legends. A systematic uncertainty of \SI{0.3}{\%} is assigned to all measurements.}
\label{fig:eff_irr}
\end{figure*}

The hit efficiency measured after irradiation as function of the bias voltage for the \SI{200}{\Omega cm} samples is shown in \fig{}~\ref{fig:eff_irr}. An efficiency over \SI{98}{\%} was obtained for all modules irradiated up to \SI{1e15}{\neqcm{}} either with protons or neutrons by applying a bias voltage larger than \SI{120}{V}. For larger irradiation fluences an efficiency lower than \SI{60}{\%} was found up to the measured voltage of \SI{160}{V} which is close to the breakdown. A larger bias voltage and lower thresholds would be necessary to operate this technology at fluences larger than \SI{1e15}{\neqcm{}}. In chips exposed to the same type of irradiation a correlation between the fluence and the hit efficiency is observed. A larger efficiency is instead measured in the chip irradiated with protons to \SI{1e15}{\neqcm{}} with respect to the one irradiated with neutrons to the same fluence despite the higher threshold of the first one. This can be explained by the larger depletion depth which is obtained with proton irradiation with respect to neutron irradiation as observed in TCT measurements due to acceptor removal effect~\cite{bern}. 
For almost all irradiated samples up to \SI{1e15}{\neqcm{}} the same efficiency as measured before irradiation is recovered with a bias voltage of \SI{150}{V}. The only exception is the H7 sensor irradiated with protons for which the measurements were limited to \SI{120}{V} by the lower breakdown voltage.

\begin{figure*}[tbph!]
\centering
	\includegraphics[width=1\textwidth]{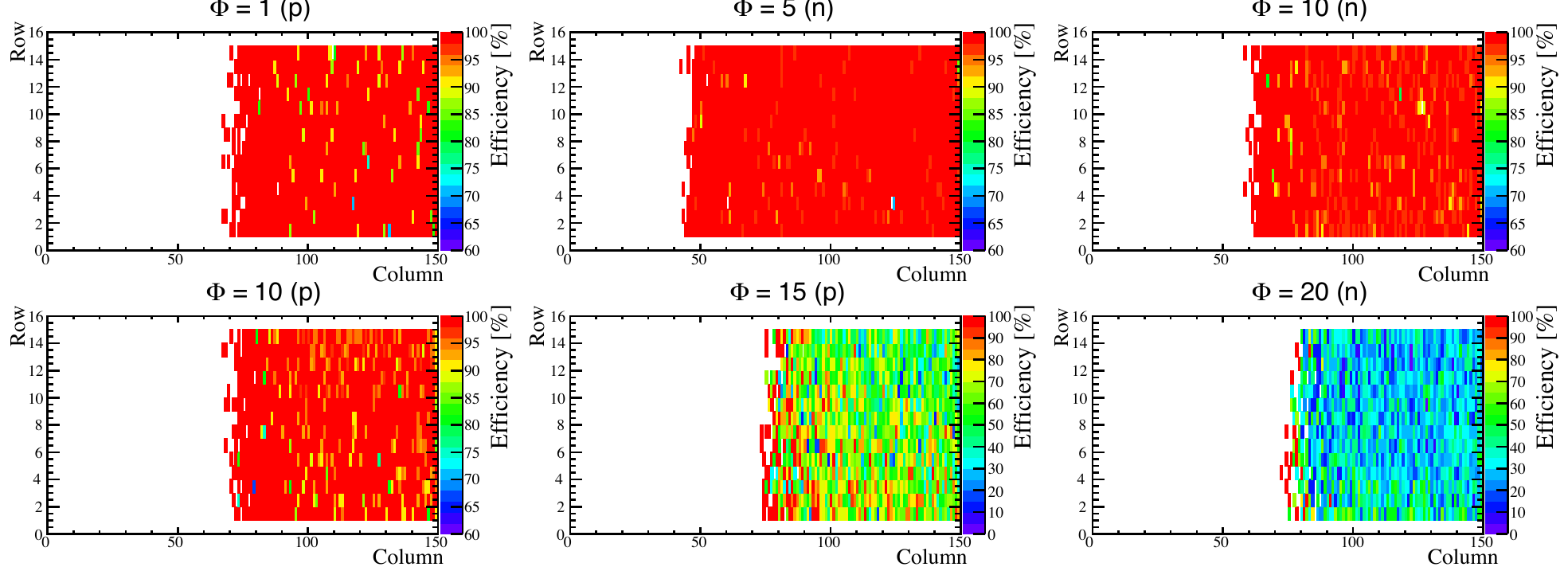}
\caption{Hit efficiency maps of \SI{200}{\Omega cm} chips after irradiation. Measurements are shown for the highest measured voltage. The fluence $\mathrm{\Phi}$ is expressed in units of \SI{1}{MeV\;\neqcm{}}. Neutron irradiation is indicated by (n) and proton irradiation by (p) in each map header. The measured area is limited on left side of the maps by the acceptance of trigger of the tracking telescope and edges are removed from the analysis to avoid smearing effects due to the telescope resolution. Note the different scale used in the efficiency for the two most irradiated devices with respect to the others.}
\label{fig:effmap_irr}
\end{figure*}

The hit efficiency over the left sub-matrix is shown in \fig{}~\ref{fig:effmap_irr} for the highest measured voltage point. As before irradiation, the efficiency is uniform over the sub-matrix surface for all irradiation types and fluences. The expected fluctuations close to the boundary of the acceptance window of the telescope trigger are observed due to the low track statistics.

\begin{figure*}[tbph!]
\centering
\subfigure[Hit efficiency vs threshold]{
	\includegraphics[width=.5\textwidth]{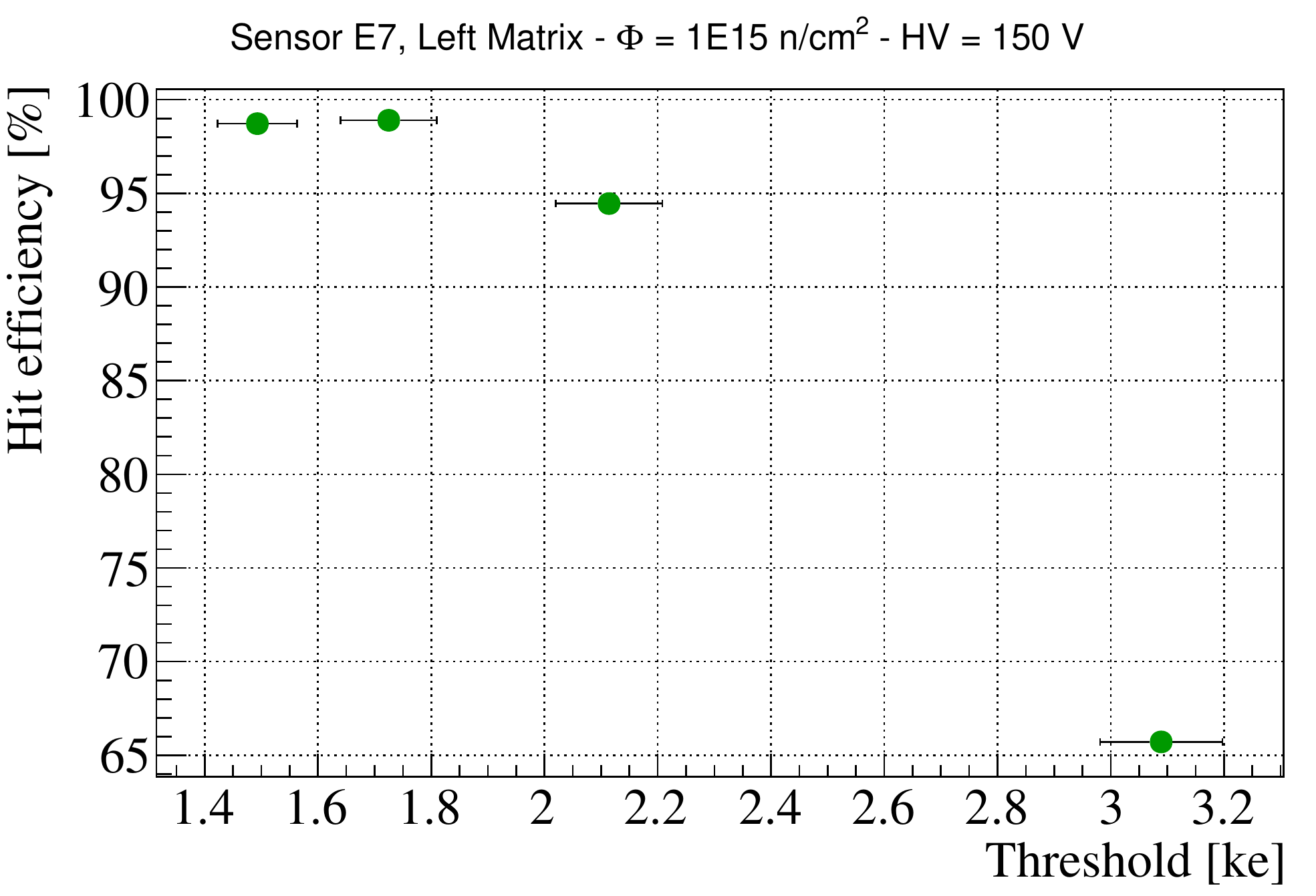}
	\label{fig:eff_irr_thr_plot}
}
\subfigure[In-pixel efficiency maps]{
	\includegraphics[width=.4\textwidth]{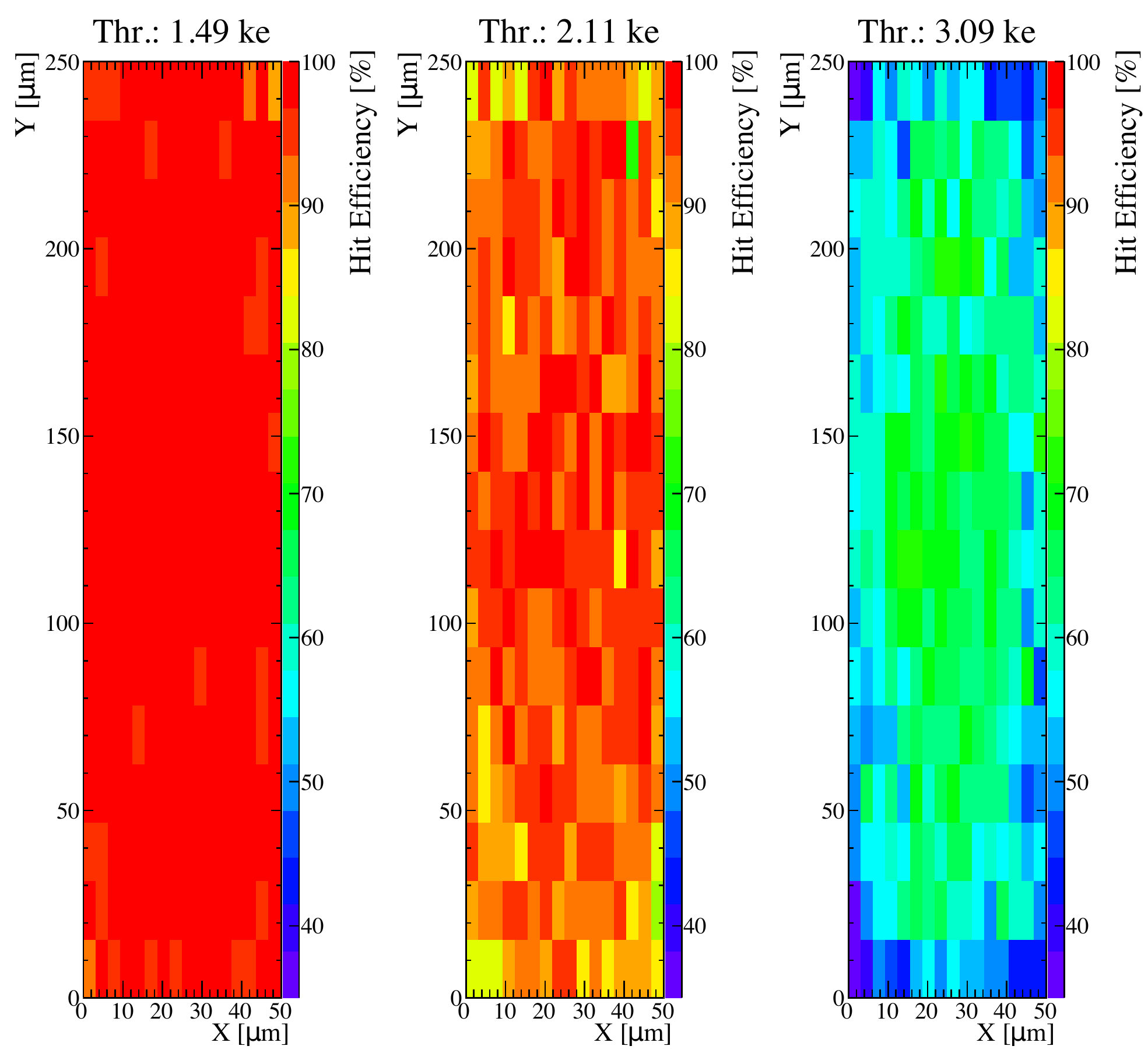}
	\label{fig:eff_irr_thr_inpix}
}
\caption{Hit detection efficiency of the CMOS left sub-matrix after neutron irradiation to a fluence of \SI{1e15}{\neqcm{}} and biased to \SI{150}{V}. In~\protect\subref{fig:eff_irr_thr_plot} the hit efficiency is shown as a function of the threshold. Error bars indicates the sigma of the gaussian fit to the correspondent threshold distribution. A systematic uncertainty of \SI{0.3}{\%} is assigned to all measurements. In~\protect\subref{fig:eff_irr_thr_inpix} the hit efficiency is shown for different threshold settings, from left to right: \SIlist{1490;2110;3090}{e}. The hit efficiency maps of the pixel cells are obtained displaying the reconstructed track impact point expressed in pixel coordinates and projecting the data for all identical structures onto the same image.}
\label{fig:eff_irr_thr}
\end{figure*}

In \fig{}~\ref{fig:eff_irr_thr} the efficiency of a sample irradiated with neutrons to \SI{1e15}{\neqcm{}} was measured as a function of the threshold while biased to \SI{150}{V}. With a threshold equal or lower than \SI{1.8}{ke} an efficiency of about \SI{99}{\%} was achieved. The efficiency gets just below \SI{95}{\%} with a threshold of about \SI{2}{ke} and then dramatically drops below \SI{70}{\%} for larger threshold values. The correspondent distribution of the hit efficiency over a pixel cell is shown in \fig{}~\ref{fig:eff_irr_thr_inpix}. When the global hit efficiency is close to \SI{99}{\%}, the hit efficiency is distributed uniformly over the pixel surface. As the threshold is increased inefficient regions are observed close to the boundaries of the cells and in particular in the corners between four pixels. This is due to charge sharing between neighbouring pixels which reduces the charge signal in each channel increasing the probability of falling below threshold.

From these results we can conclude that a bias voltage equal or larger than \SI{130}{V} and a maximum threshold of \SI{1.8}{ke} are necessary to operate \SI{200}{\Omega cm} samples with an efficiency larger than \SI{97}{\%} up to irradiation fluences of \SI{1e15}{\neqcm{}}.


\section{Noise occupancy and power consumption}\label{sec:noise_occ}
In order to fully evaluate the performance of the chip the results of the hit efficiency need to be evaluated together with the noise occupancy. For the ATLAS experiment a noise occupancy per pixel of less than \SI{e-6}{} hits in \SI{25}{ns}, i.e. a LHC bunch crossing, is required.

The noise occupancy of the left sub-matrix was measured for irradiated and non-irradiated \SI{200}{\Omega cm} chips in a climate chamber at stable temperatures: Non-irradiated devices were kept at \SI{20}{\celsius}, while irradiated devices were cooled down to \SI{-35}{\celsius}. This is estimated to be equivalent to a temperature on chip, when configured, of approximately \SI{30}{\celsius} and \SI{-25}{\celsius}, respectively. All chips were operated at the voltage for which the highest hit efficiency was measured and in the same conditions as they were operated at beam tests. An acquisition window of \SI{5}{minutes} was used in absence of radioactive sources to integrate the noise hits. Given the small dimensions of the matrix the cosmic muon background is considered negligible. 

\begin{figure*}[tbph!]
\centering
	\includegraphics[width=.6\textwidth]{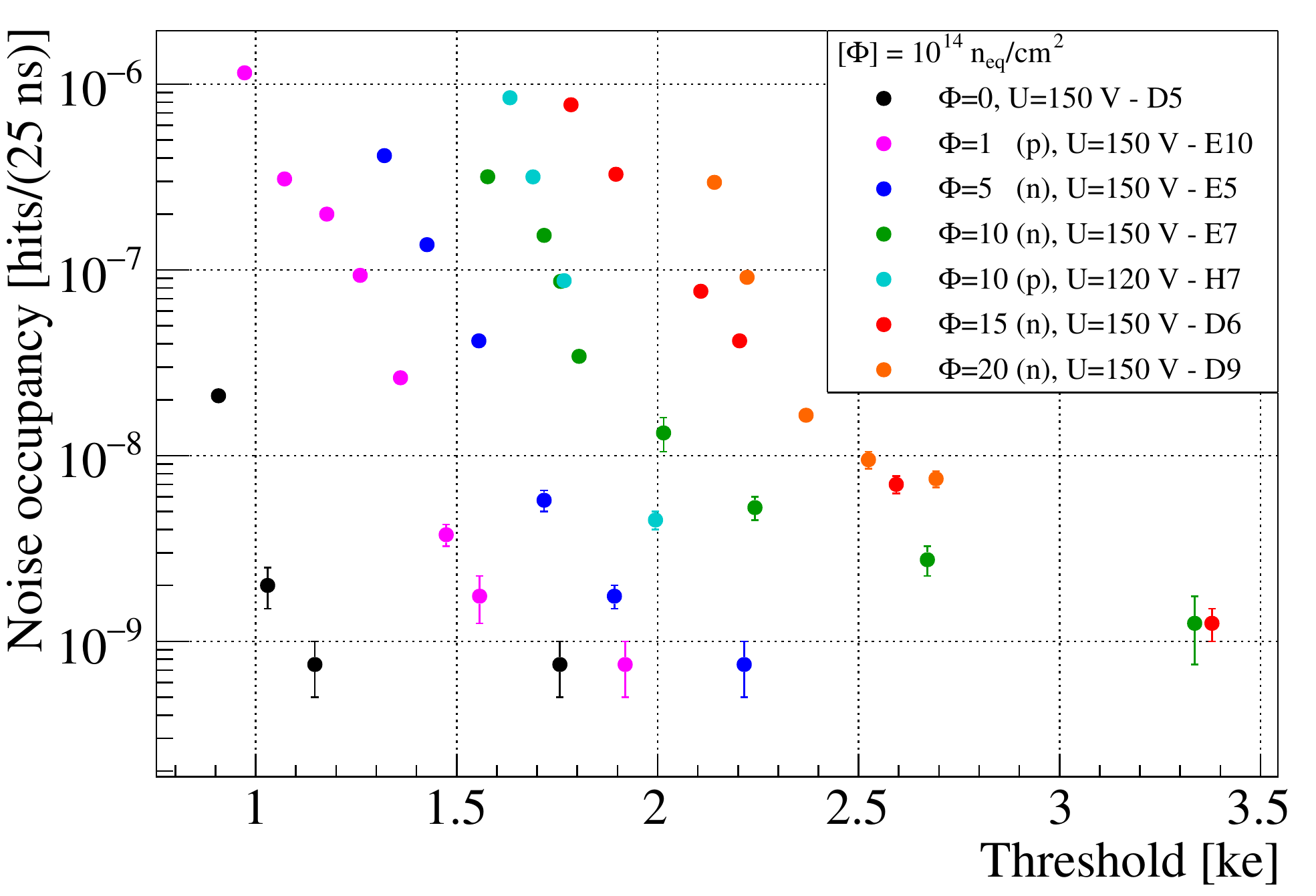}
\caption{Noise occupancy in one LHC bunch crossing of \SI{25}{ns} as a function of the mean threshold for the left sub-matrix of the CMOS monolithic matrix.}
\label{fig:noise_occ}
\end{figure*}

         Noise occupancy results as a function of the threshold are showed in \fig{}~\ref{fig:noise_occ}. Before irradiation the noise occupancy in the full left sub-matrix reaches a maximum of \SI{2e-8}{} at the lowest measured threshold of \SI{900}{e}, otherwise it falls below \SI{e-9}{} with thresholds higher than \SI{1100}{e}. After irradiation the noise occupancy can be still kept below \SI{e-6}{} in the case of neutron irradiations even with thresholds between \SIlist{1300;1800}{e} with which a hit efficiency of about \SI{99}{\%} was achieved at beam tests. This correspond to a noise occupancy per pixel of about \SI{4e-10}, well below the ATLAS requirements.

A power consumption of about \SI{97}{\mu W} per pixel was measured using the DAC settings with which the chips were operated at the beam test. This includes a contribution of \SI{91}{\mu W} per pixel from the digital part which is powered with \SI{3.3}{V}, and \SI{6}{\mu W} per pixel from the analog part which is powered with \SI{2}{V}. These results are consistent with the simulations published in ref.~\cite{h35demo}. No significant variation of the power consumption of the chip was observed after irradiation apart from the need of rising the digital voltage to avoid the mentioned crosstalk effect discussed in \sect{}~\ref{sec:irrad}.

\section{Conclusions and outlook}\label{sec:conclusions}
Very first results of the characterisation of the H35DEMO chip have been presented demonstrating the feasibility of producing large monolithic devices using a cost-effective CMOS process to fulfil the requirements in terms of radiation hardness for the ATLAS experiment at HL-LHC. The monolithic CMOS matrix of the H35DEMO was operated for minimum ionising particle detection using the DAQ system developed at IFAE.
Before irradiation a comparison of the performance of different resistivities measured at beam tests showed best results for \SIlist{80;200}{\Omega cm} with respect to the usual \SI{20}{\Omega cm} resistivity substrates commonly used in the industrial CMOS process. Devices with higher resistivity substrates were able to reach a detection efficiency around \SI{99}{\%} with a bias voltages lower than \SI{80}{V}.

After irradiation several problems have been identified in the monolithic part of this first large area prototype which need to be considered in the designs of future chips. Nevertheless, a characterisation of detection performance was possible up the radiation fluences expected for the outermost pixel layer of the ATLAS experiment at HL-LHC. The results demonstrated the radiation hardness of this large electrode design after neutron irradiation up to a fluence of \SI{1e15}{\neqcm{}} showing an hit efficiency of \SI{99}{\%} at \SI{150}{V} with an occupancy of less than \SI{1e-6}{} noise hits in a LHC bunch crossing of \SI{25}{ns}. A hit efficiency of more than \SI{98}{\%} was also measured for proton irradiated chips with \SI{130}{V} after a particle fluence of \SI{1e15}{\neqcm{}}. 

The performance of this technology could be taken farther by improving the tuning capabilities to reduce the threshold dispersion, and implement the possibility of effectively masking single pixels. Moreover, the performance of this chip is also limited by the H35 process which was chosen to contain the costs and eventually investigate this technology for less demanding applications in terms of radiation hardness. In particular, by moving to a CMOS process in \SI{180}{nm} that allows to fit more functionalities within a smaller pixel area, the radiation hardness of the digital part will be improved and at the same time the capacitance of the pixel reduced. Moreover, the power dissipation will also improve given the lower bias voltages required for the electronics in this technology with respect the H35. The H18 technology is not only available at AMS, but also in other foundries, such as TSI~\cite{tsiweb} in the United States, giving the flexibility of moving the process to another foundry in case of issues and in view of large productions.	

The ATLASPix, a series of large area HV-CMOS chips has been already designed in CMOS \SI{180}{nm} H18 AMS technology and the first prototype, the ATLASPix1, has been already produced and successfully tested before irradiation~\cite{atlaspix1}. The ATLASPix2 has been also submitted for production at TSI in June 2018. This new generation of monolithic prototypes for ATLAS aims at moving towards a final demonstrator by consolidating the radiation hardness features of this design and including all the functionalities required by the ATLAS experiment.
In particular it will have to investigate the possibility of matching the requirements in terms of timing and in-time efficiency which were not met by previous prototypes as shown by the measurements preformed on CCPDs~\cite{h18irr,h35ccpd}.

\acknowledgments
This work was partially funded by: the Generalitat de Catalunya (AGAUR 2014 SGR 1177), the MINECO, Spanish Government, under grants FPA2015-69260-C3-2-R, FPA2015-69260-C3-3-R (co-financed with the European Union's FEDER funds) and SEV-2012-0234 (Severo Ochoa excellence programme), under the Juan de la Cierva programme; the European Union's Horizon 2020 Research and Innovation programme under Grant Agreement no. 654168; and the PhD fellowship program of La Obra Social La Caixa-Severo Ochoa.
The authors would like to thank A.~Dierlamm and F.~B\"ogelspacher (KIT) as well as V.~Cindro and I.~Mandic (JSI) for the excellent support for the irradiations, and S.~K\"uhn and S.~Dungs for the availability and support with the x-ray fluorescence setup at CERN.
\FloatBarrier

\end{document}